\documentclass[pdflatex,sn-mathphys-num]{sn-jnl}% Math and Physical Sciences Numbered Reference Style
\usepackage{graphicx}%
\usepackage{multirow}%
\usepackage{amsmath,amssymb,amsfonts}%
\usepackage{amsthm}%
\usepackage{mathrsfs}%
\usepackage[title]{appendix}%
\usepackage{xcolor}%
\usepackage{textcomp}%
\usepackage{manyfoot}%
\usepackage{booktabs}%
\usepackage{algorithm}%
\usepackage{algorithmicx}%
\usepackage{algpseudocode}%
\usepackage{listings}%
\theoremstyle{thmstyleone}%
\theoremstyle{thmstyletwo}%

\theoremstyle{thmstylethree}%

\begin{document}

\title[DebrisWatch II]{DebrisWatch II: Digging deeper for geosynchronous debris}


\author*[1,2,3]{\fnm{James A.} \sur{Blake}}\email{j.blake.1@warwick.ac.uk}

\author[1,2,3]{\fnm{Benjamin F.} \sur{Cooke}}%\email{iiauthor@gmail.com}

\author[4]{\fnm{Cristina} \sur{Paragini}}%\email{iiiauthor@gmail.com}

\author[5]{\fnm{William} \sur{Feline}}
\author[6]{\fnm{Christopher A.} \sur{Onken}}
\author[1,2,3]{\fnm{Don} \sur{Pollacco}}
\author[7]{\fnm{Grant} \sur{Privett}}
\author[8]{\fnm{Toshifumi} \sur{Yanagisawa}}
\author[1,2]{\fnm{Robert} \sur{Airey}}
\author[1,2,3]{\fnm{Ioannis} \sur{Apergis}}
\author[4]{\fnm{Roberto} \sur{Armellin}}
\author[9]{\fnm{Lily} \sur{Beesley}}
\author[1,2]{\fnm{Paul} \sur{Chote}}
\author[1]{\fnm{Anna-Maria} \sur{Cutolo}}
\author[10]{\fnm{Stuart} \sur{Eves}}
\author[11]{\fnm{Tomoko} \sur{Fujiwara}}
\author[11]{\fnm{Daisuke} \sur{Kuroda}}
\author[1,3]{\fnm{Isobel S.} \sur{Lockley}}
\author[1,2,7]{\fnm{Alexander} \sur{MacManus}}
\author[1,2,3]{\fnm{James} \sur{McCormac}}
\author[1,3]{\fnm{Morgan A.} \sur{Mitchell}}
\author[11]{\fnm{Tokuhiro} \sur{Nimura}}
\author[11]{\fnm{Kota} \sur{Nishiyama}}
\author[11]{\fnm{Shin-ichiro} \sur{Okumura}}
\author[12]{\fnm{Thomas} \sur{Schildknecht}}
\author[1,2]{\fnm{Billy} \sur{Shrive}}
\author[11]{\fnm{Seitaro} \sur{Urakawa}}
\author[1,2,3]{\fnm{Dimitri} \sur{Veras}}
%\author[1,2,13]{\fnm{Han} \sur{Wang}}
\author[1,2]{\fnm{Phineas} \sur{Whitlock}}
\author[6]{\fnm{Christian} \sur{Wolf}}

\affil*[1]{\orgdiv{Department of Physics}, \orgname{University of Warwick}, \orgaddress{\street{Gibbet Hill Road}, \city{Coventry}, \postcode{CV4 7AL}, \country{United Kingdom}}}

\affil*[2]{\orgdiv{Centre for Space Domain Awareness}, \orgname{University of Warwick}, \orgaddress{\street{Gibbet Hill Road}, \city{Coventry}, \postcode{CV4 7AL}, \country{United Kingdom}}}

\affil[3]{\orgdiv{Centre for Exoplanets and Habitability}, \orgname{University of Warwick}, \orgaddress{\street{Gibbet Hill Road}, \city{Coventry}, \postcode{CV4 7AL}, \country{United Kingdom}}}

\affil[4]{\orgdiv{Te Pūnaha Ātea - Space Institute}, \orgname{University of Auckland}, \orgaddress{\street{Princes Street}, \city{Auckland}, \postcode{1010}, \country{New Zealand}}}

\affil[5]{\orgname{Defence Science \& Technology Laboratory}, \orgaddress{\street{Portsdown West}, \city{Fareham}, \postcode{PO17 6AD}, \country{United Kingdom}}}

\affil[6]{\orgdiv{Research School of Astronomy and Astrophysics}, \orgname{Australian National University}, \orgaddress{\street{Bart Bok Road}, \city{Canberra}, \postcode{ACT 2611}, \country{Australia}}}

\affil[7]{\orgname{Defence Science \& Technology Laboratory}, \orgaddress{\street{Porton Down}, \city{Salisbury}, \postcode{SP4 0JQ}, \country{United Kingdom}}}

\affil[8]{\orgname{Japan Aerospace Exploration Agency}, \orgaddress{\street{7-44-1 Jindaiji Higashi-machi}, \city{Chofu-shi}, \postcode{Tokyo 182-8522}, \country{Japan}}}

\affil[9]{\orgdiv{School of Metallurgy and Materials}, \orgname{University of Birmingham}, \orgaddress{\street{Elms Road}, \city{Birmingham}, \postcode{B15 2SE}, \country{United Kingdom}}}

\affil[10]{\orgname{SJE Space Ltd}, \orgaddress{\street{Clayhill Road}, \city{Reading}, \postcode{RG7 3HB}, \country{United Kingdom}}}

\affil[11]{\orgname{JAXA Bisei Space Guard Center}, \orgaddress{\street{1716-3 Biseicho Okura}, \city{Ibara}, \postcode{Okayama 714-1411}, \country{Japan}}}

\affil[12]{\orgname{Space Eye Observatory}, \orgaddress{\street{Uecht 53a}, \postcode{3087 Niedermuhlern}, \country{Switzerland}}}

%\affil[13]{\orgdiv{School of Aeronautics and Astronautics}, \orgname{Sun Yat-sen University}, \orgaddress{\city{Shenzhen}, \country{China}}}


\abstract{
\unboldmath
The geosynchronous (GSO) debris environment is continually evolving. 
Regular monitoring of the region is consequently of great importance, though the trade-off between coverage and sensitivity makes this challenging for the population of optically faint debris, where collecting area becomes a pivotal factor.
Surveys conducted with large-aperture telescopes have provided crucial insights into the nature of this largely uncharacterised population.
In this paper, we revisit a survey conducted with the 2.54\,m Isaac Newton Telescope (INT), presenting an overhaul of the astrometric calibration and object detection stages of the original analysis pipeline.
We apply a blind stacking technique to boost target recovery, unearthing 25 tracklets previously missed by single-frame extraction methods, and pushing the sensitivity limit fainter by 1 magnitude.
The same algorithm is applied to a contemporaneous dataset, captured with a 36\,cm astrograph, enabling performance benchmarking through the attempted recovery of INT detections from commercial-off-the-shelf observations.
We achieve sub-arcsecond astrometric accuracy through a combination of improved star trail centroiding and iterative distortion fitting, allowing short arc initial orbit solutions to be obtained.
High-cadence light curves extracted for trailing detections indicate that faint fragments are proportionally more variable than bright derelicts, with many exhibiting photometric signatures of rapid tumbling, often straddling the image noise floor.
Lastly, we present preliminary findings from a follow-up multi-national observation campaign, utilising telescopes in Australia, Japan and La Palma.
As space traffic management concerns begin to extend beyond GSO altitudes, scientifically-driven surveys of high-altitude orbits have an important role to play in characterising the faint debris environment.
}




\maketitle

\section{Introduction}
\label{sec:introduction}

%The Introduction section, of referenced text \cite{bib1} expands on the background of the work (some overlap with the Abstract is acceptable). The introduction should not include subheadings.

%Springer Nature does not impose a strict layout as standard however authors are advised to check the individual requirements for the journal they are planning to submit to as there may be journal-level preferences. When preparing your text please also be aware that some stylistic choices are not supported in full text XML (publication version), including coloured font. These will not be replicated in the typeset article if it is accepted.

The Space Age is undoubtedly accelerating into a new era, characterised by the privatisation of space flight, increased access to space, and soaring launch rates to low Earth orbit, as large constellations of satellites begin to take shape~\citep{blake2022looking,lawrence2022case,mcdowell2020low}. 
Amidst this surge of activity close to home, it can be easy to overlook the higher-altitude orbits that have long supported a variety of broadcasting, surveillance, communications, and meteorological services.
Nevertheless, the geosynchronous Earth orbit (GSO) region will continue to play a crucial role. 
Emerging business models are enhancing the affordability and flexibility of GSO operations, while a diversification of platforms and missions is fostering novel applications. 
An increasing number of multi-orbit solutions are leveraging the fixed-positioning and long-lifetime capabilities of geostationary satellites.

\begin{figure}[tbp]
\centering
\includegraphics[width=\textwidth]{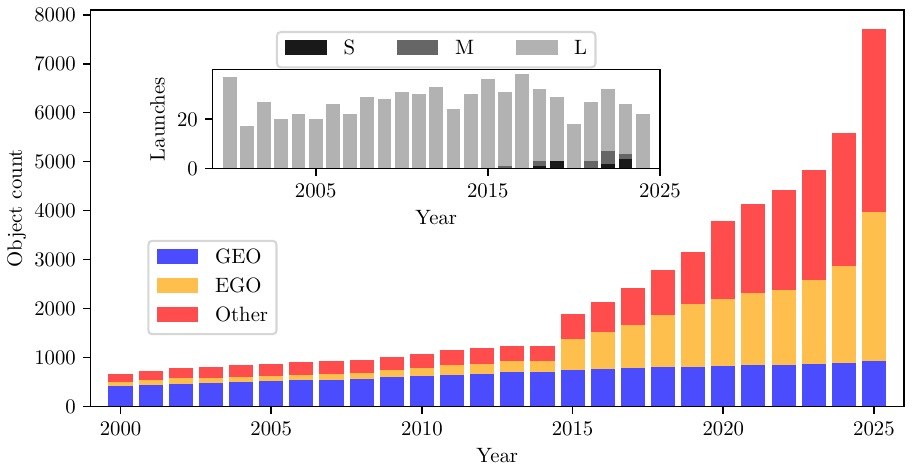}
\caption{Recent evolution of the GSO region, sourced from the latest edition of the European Space Agency's Annual Space Environment Report~\citep{esa2025space}.
The main panel shows the number of RSOs penetrating the GSO Protected Region, defined by the ranges $h\in\left[35586,35986\right]$\,km and $\delta\in\left[-15,15\right]^\circ$, where $h$ is altitude and $\delta$ is declination.
The RSOs are categorised according to their orbit type: `GEO' (geostationary) with inclination $i\in\left[0,25\right]^\circ$, perigee height $h_p\in\left[35586,35986\right]$\,km, and apogee height $h_a\in\left[35586,35986\right]$\,km; `EGO' (extended geostationary) with semi-major axis $a\in\left[37948,46380\right]$\,km, eccentricity $e\in\left[0.00,0.25\right]$, and $i\in\left[0,25\right]^\circ$; and `Other', encompassing all other types, including GEO transfer and highly eccentric orbits.
Inset, GEO launch traffic is provided, with payloads categorised according to their mass $m$: `S' (small), with $10<m\leq100$\,kg; `M' (medium), with $100<m\leq1000$\,kg; and `L' (large), with $m>1000$\,kg.}
\label{fig:geo-population}
\end{figure}

Furthermore, the GSO environment has served as an effective testing ground for a variety of circular economy initiatives, most notably Northrop Grumman’s Mission Extension Vehicle demonstrations~\citep{george2021phantom} and China’s tugging of a defunct navigation satellite from the geostationary belt to a supersynchronous graveyard orbit using SJ-21~\citep{weeden2022chinese}.
Missions involving rendezvous and proximity operations, such as these, require high levels of situational awareness to maintain custody of the spacecraft involved and to prevent accidents from taking place.

Debris generation is especially problematic at GSO altitudes.
Contrary to low-altitude orbits, where atmospheric drag acts as a natural removal mechanism, no equivalent process exists to clear resident space objects (RSOs) from the GSO region.
The rapid growth of the commercial space domain awareness sector has significantly improved the timeliness and persistence of GSO monitoring, enabling the swift detection and characterisation of anomalies and fragmentation events~\citep{ingram2022exoalert}.
Most recently, the breakup of Intelsat 33E in October 2024 has added 36 fragments to the publicly available Space-Track catalogue, as of the time of writing~\citep{spacetrack,nasa2025quaterly29-1}.
Many tens to hundreds of additional fragments continue to be tracked by responding surveillance networks~\citep[see, e.g.,][]{pavy2025rapid,pirovano2025northstar}, but maintaining custody of optically faint debris in high-altitude orbits remains a significant challenge.
Furthermore, breakup events can generate many thousands of fragments that are too faint to be detected by ground-based instrumentation, as demonstrated by laboratory experiments~\citep{cowardin2023updates} and reflected in the space debris environment models that draw from them~\citep{braun2021recent}.
An overview of recent GSO anomalies and fragmentation events is provided in Table~\ref{tab:fragmentation-events}, collectively accounting for large parts of the steep increase in the number of trackable high-altitude RSOs that can be seen in Fig.~\ref{fig:geo-population}.
It is important to note that the rise in object count also reflects improvements in surveillance capabilities that have come about over the past decade, and many of the fragments discovered during this period will have originated from events that took place long before.

\begin{table}[tbp]
	\centering
	\caption{Notable (well-reported) anomalies and fragmentation events in high-altitude orbits that have taken place in the past decade. 
    Information sourced from a mixture of academic/conference literature and newsletters, as indicated in the final column, alongside the \textit{History of On-orbit Satellite Fragmentations} database~\cite{anz2023history}. 
    Two counts are provided for the number of trackable fragments generated by each event: the number of associated entries that appear in the publicly available Space-Track catalogue~\citep{spacetrack}; and, in parentheses, the number of fragments claimed to have been initially identified by responding survey campaigns.
    Note that these inventories are limited by the sensitivities of the instrumentation employed, thus they will severely underestimate the true number of fragments generated across the full range of fragment sizes.}
	\label{tab:fragmentation-events}
	\begin{tabular}{@{}llllllll@{}}
        \toprule
            Name & INTLDES & Event date & Apo/Perigee [km] & Incl. [$^\circ$] & Fragments & Ref. \\
        \midrule
            %2015-075B & 16-Jan-2016 & 35777/33426 &  & 7 \\
            AMC-9 & 2003-024A & 17-Jun-2017 & 35798/35774 & 0.0174 & 0 (13+) & \cite{cunio2017photometric} \\
            Telkom 1 & 1999-042A & 25-Aug-2017 & 35793/35781 & 0.0112 & 0 (10+) & \cite{cunio2017photometric} \\
            Titan 3C Tra. & 1969-013B & 28-Feb-2018 & 37257/35886 & 6.2 & 18 (150+) & \cite{schildknecht2019esa} \\
            Atlas 5 Cent. & 2014-055B & 30-Aug-2018 & 35090/8112 & 22.3 & 109 (590+) & \cite{schildknecht2019esa} \\
            Atlas 5 Cent. & 2009-047B & 24-Mar-2019 & 34565/6810 & 23.3 & 1 (250+) & \cite{schildknecht2019esa} \\
            Atlas 5 Cent. & 2018-079B & 6-Apr-2019 & 35092/8526 & 12.2 & 214 (250+) & \cite{schildknecht2019esa} \\
            Intelsat 29e & 2016-004A & 7-Apr-2019 & 35707/35675 & 0.09 & 2 (10+) & \cite{cunio2019remote} \\
            Atlas 5 Cent. & 2018-022B & 6-Sep-2024 & 34949/7622 & 9.3 & 0 (10+) & \cite{nasa2025quaterly29-1} \\
            Intelsat 33e & 2016-053B & 19-Oct-2024 & 35797/35773 & 0.037 & 36 (500+) & \cite{pirovano2025northstar} \\
        \botrule
	\end{tabular}
\end{table}

Observations of the GSO region are typically carried out using optical telescopes, as their sensitivity drops off more gradually than that of alternative active sensing techniques, such as radar.
Early attempts to probe the population of small GSO debris were made by NASA using the Charge-Coupled Device (CCD) Debris Telescope (CDT), a portable 32\,cm Schmidt telescope initially based in Maui, Hawaii, and subsequently at Cloudcroft, New Mexico~\citep{jarvis2001observations}.
Despite the instrument's small format and large pixel scale, the CDT was able to detect many $\sim1$\,m RSOs that were not previously catalogued.
These efforts were followed by two long-lasting and highly influential programmes: the NASA-funded Michigan Orbital DEbris Survey Telescope (MODEST)~\citep{seitzer2004modest}; and ESA's 1\,m telescope campaigns~\citep{schildknecht2024recent}.
MODEST is a 0.6\,m Schmidt telescope located at the Cerro Tololo Inter-American Observatory in Chile.
From 2001 to 2014, MODEST was used to carry out extensive observations of the GSO region, providing coverage of most orbital slots over the continental US.
The system achieved a relatively wide field of view (FOV) of $1.3^\circ\times1.3^\circ$ and a limiting magnitude of roughly 17.5 in the Cousins $R$ band.
Its successor, the 1.3\,m Eugene Stansbery-Meter Class Autonomous Telescope, has recently been deployed on Ascension Island~\citep{cruz2025detection}.
The ESA 1\,m telescope is installed at the Optical Ground Station at Teide Observatory in Tenerife, Canary Islands.
Since 2001, the system has been used by the Astronomical Institute of the University of Bern (AIUB) to conduct regular surveys of the GSO region and other high-altitude orbital regimes, often supported by instruments at AIUB's Zimmerwald Observatory in Switzerland~\citep[see, e.g.,][]{silha2017optical,schildknecht2008properties}.
Compared to MODEST, the ESA 1\,m has a smaller FOV of $0.7^\circ\times0.7^\circ$, but its wider collecting area enables the detection of sources fainter than 19$^\text{th}$ magnitude, corresponding to RSOs $10-20$\,cm in size, assuming a Lambertian sphere with an albedo of 0.1~\cite{africano2005understanding}.

Both the MODEST and ESA 1\,m campaigns served to highlight a bimodal brightness distribution, with known (catalogued) satellites and rocket bodies clustered around 12$^\text{th}$ magnitude and an abundance of uncatalogued RSOs fainter than 15$^\text{th}$ magnitude.
The bright-end drop-off and the dearth of objects between the two peaks are characteristics of the GSO population, while the faint-end drop-off is due to the sensitivity limit of the instrument.
It is also important to note that the brightness of RSOs is strongly dependent on the illumination geometry between the Sun, the target, and the observer at the time of observation~\citep[see, e.g.,][]{airey2025comprehensive}.
As surveys typically report an average brightness across an observation window (or multiple windows), the characteristics of the brightness distribution will vary according to the observational strategy employed.

Surveys attempting to detect faint GSO debris usually track the telescope and detector at the same rate as geostationary satellites.
However, target RSOs are moving relative to this rate, as uncontrolled debris will drift under the influence of the natural perturbative forces that dominate in the GSO region~\citep{mcknight2013new}.
Increasing the exposure time to boost the signal-to-noise ratio (SNR) and push the limiting magnitude fainter is consequently not an option, as longer exposure times will spread the target signal across a trail of pixels, while the background noise accumulates proportionally to the trail area.
It proves necessary, therefore, to appeal to larger-aperture telescopes for their superior collecting areas.

Large-aperture nodes of the International Scientific Optical Network (ISON), a non-government project led by the Keldysh Institute of Applied Mathematics, Russia, have been used to detect and continuously track faint debris in high-altitude orbits~\citep{molotov2009faint}, most notably the narrow-field 1.25\,m and 2.6\,m telescopes at the Crimean Astrophysical Observatory in Nauchny, with limiting magnitudes of 19 and 20, respectively.
Working in collaboration with AIUB, ISON telescopes conducted follow-up observations of high-altitude RSOs with high area-to-mass ratios, extending coverage and enabling a catalogue of these objects to be built and maintained~\citep{schildknecht2012long}.
The ISON project continues to conduct research into a variety of orbital regimes with a geographically diverse network of over 50 optical telescopes~\citep{molotov2024international}.
Bolden et al. \cite{bolden2011panstarrs} carried out observations of faint GSO debris using the 1.8\,m Pan-STARRS PS1 telescope on Haleakala, Maui, achieving a survey depth of $V\sim21$. 
Interestingly, the Pan-STARRS survey observed a sharp increase in the number of very faint RSOs (fainter than $V\sim19.5$) before the instrumental drop-off in sensitivity.
This is unsurprising, as the instrument is well-situated to probe the surroundings of the geopotential well at 105$^\circ$\,W, where many small RSOs originating from fragmentation events will be oscillating about the well centre if captured, and such events will typically generate many more small fragments than large ones~\citep{murray2019analysis}.  
That said, the authors exercise caution, highlighting partial streak detection as a possible source of brightness underestimation. 
The 6.5\,m Magellan Telescope, `Walter Baade', at the Las Campanas Observatory, Chile, has been used to conduct a number of `spot' surveys targeting known fragmentation events~\citep{seitzer2017search}.
Despite its relatively narrow 0.2 square degree FOV, the telescope detected a significant number of RSOs with brightness $18<R<20$, many of which were glinting rapidly.

\begin{table}[tbp]
	\centering
	\caption{Specifications for the instruments employed for the 2018 survey. 
    Note that the quoted field of view for the INT is lower than its achievable $34\times34$ square arcminutes when paired with the WFC, owing to severe pickup noise that rendered one of the mosaic's CCD chips unusable.}
	\label{tab:instrument-specifications}
	\begin{tabular}{@{}lll@{}}
        \toprule
             & INT (WFC) & RASA \\
        \midrule
            Aperture [m] & 2.54 & 0.36 \\
            CCD(s) [px] & 4~$\times$~2k~$\times$~4k & 8k~$\times$~6k \\
            FOV & $33'\times22'$ & $3.6^\circ\times2.7^\circ$ \\
            Pixel scale [$''$px$^{-1}$] & 0.66\textsuperscript{a} & 1.57 \\
            Readout time [s] & 25 & 4 \\
            Filter & Harris $V$ & None \\
        \botrule
	\end{tabular}
    \footnotetext{\textsuperscript{a} The quoted pixel scale for the INT is with $2\times2$ binning in place.}
\end{table}

In this paper, we revisit a survey of faint GSO debris conducted with the Isaac Newton Telescope~\citep{blake2021debriswatch}.
We provide a brief overview of the original survey in Section~\ref{sec:instrumentation-strategy}, introducing the instrumentation and observational strategy employed.
In Sections~\ref{sec:astrometric-calibration} and \ref{sec:blind-stacking}, we outline improvements made to the astrometric calibration and RSO detection stages of the survey analysis pipeline, extending the discussions presented in \cite{blake2025improved}.
Exploiting the improved astrometry and photometry, we perform a preliminary orbit analysis for the very short INT arcs in Section~\ref{sec:short-arc-orbit-determination}, before analysing high-cadence light curves extracted for trailing detections in Section~\ref{sec:brightness-variability}.
Lastly, we present early findings from a follow-up multinational observing campaign in Section~\ref{sec:future-work}, before exploring avenues for future work and concluding the paper in Section~\ref{sec:conclusions}.

\section{Instrumentation and strategy}
\label{sec:instrumentation-strategy}

The data used in this study were obtained with the 2.54\,m Isaac Newton Telescope (INT) and a commercial-off-the-shelf (COTS) robotic instrument (hereafter, RASA), both located at the Roque de los Muchachos Observatory in La Palma, Canary Islands, as part of a coordinated optical survey of the GSO region.
For ease of reference, we provide a brief overview of the original survey here, referring the interested reader to Blake et al.~\cite{blake2021debriswatch} for further details.

\begin{figure}[tbp]
\centering
\includegraphics[width=\textwidth]{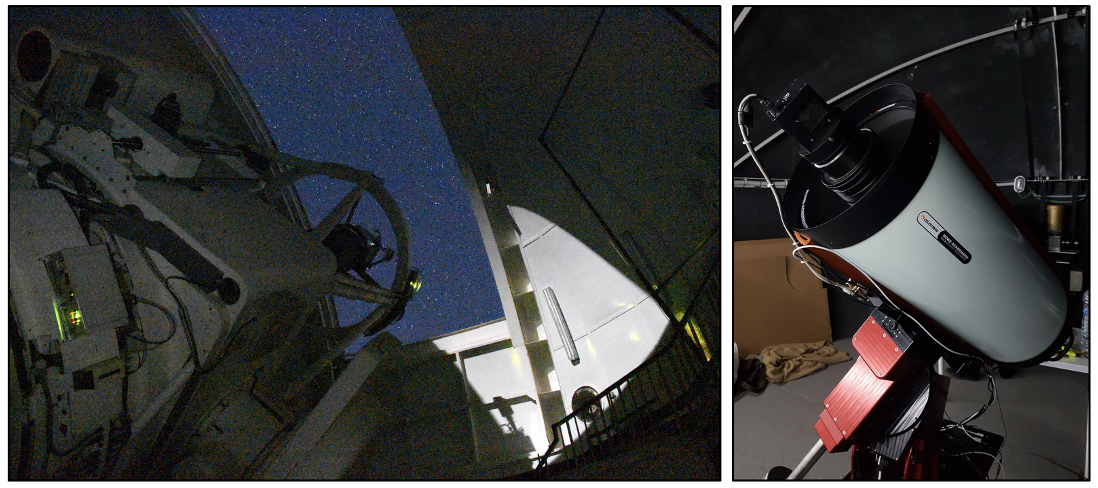}
\caption{The 2.54\,m Isaac Newton Telescope, with prime-focus Wide Field Camera, alongside the 36\,cm robotic astrograph used for the survey.}
\label{fig:int-rasa}
\end{figure}

The survey spanned eight nights of dark-grey time from 2–9 September 2018.
Observations with the INT were made using the prime-focus Wide Field Camera (WFC), a mosaic comprising four thinned charge-coupled devices (CCDs), while the RASA paired a 36\,cm Rowe-Ackermann Schmidt astrograph with an FLI ML50100 CCD camera. 
The two instruments, shown in Fig.~\ref{fig:int-rasa}, observed the same regions of sky contemporaneously. Relevant specifications are provided in Table~\ref{tab:instrument-specifications}.

\begin{figure}[tbp]
\centering
\includegraphics[width=\textwidth]{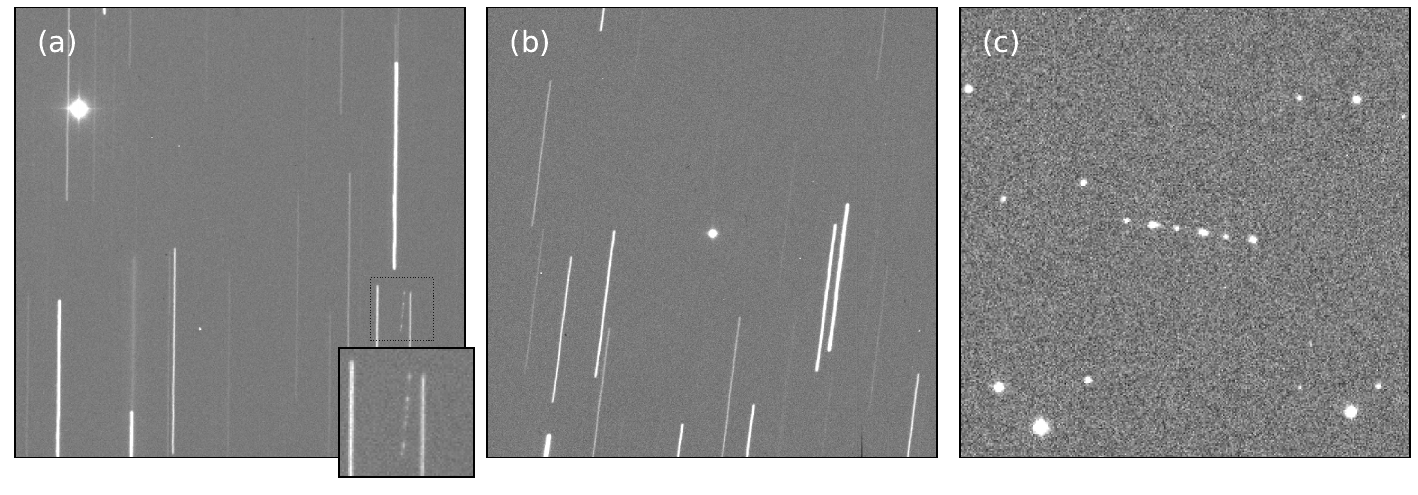}
\caption{Observational strategies for monitoring objects in high-altitude orbits, employing different tracking rates.
(a) An exemplar frame from the survey in question, acquired with the INT `stopped'.
Two RSO detections are contained within the window: a point-like (though saturated) geostationary satellite in the top left corner; and a faint (glinting) trail associated with an uncatalogued fragment of GSO debris, expanded for clarity.
(b) Separate to the main survey, differential tracking rates were applied to the INT to match the drifting motion of a Titan 3C rocket body (NORAD 5589) in a graveyard orbit, integrating photons of reflected light from the target onto the same pixels for the duration of the exposure.
(c) A sidereally tracked image of the Sirius 3 satellite (NORAD 25492), taken with the RASA instrument, also separate from the main survey.
Short-period glints are apparent along the target's otherwise faint trail.
Note that stars exhibit predictable and uniform behaviour in each mode of operation, though their morphologies and orientations can vary greatly depending on the chosen strategy.
Adapted from~\cite{blake2021optical}.}
\label{fig:observational-strategies}
\end{figure}

Images were acquired with the telescopes fixed (i.e., stationary with respect to an Earth-fixed frame of reference) to ensure that photons from potential GSO candidates would integrate over fewer pixels during the exposure.
In this mode of operation, station-kept geostationary satellites will image as point sources (blurred according to the point spread function of the optical system), while other GSO targets will manifest as short trails, as illustrated in panel (a) of Fig.~\ref{fig:observational-strategies}.
Background stars will form longer trails aligned with the longitudinal axis, with an extent determined by the pixel scale and exposure time.
An exposure time of 10\,s was chosen to provide a reasonable balance between minimising stellar streak coverage of the CCDs and maximising the observational duty cycle.

Target fields were selected based on their proximity to the penumbral boundary of the Earth's shadow projection at geostationary altitude in order to minimise the solar phase angle (observer-target-Sun) and thus maximise the apparent brightness of candidate RSOs, while at all times avoiding eclipse by the shadow.
Each pointing of the INT comprised a sequence of seven 10\,s exposures, with a readout time of 25\,s separating the exposures.
The telescope pointing was updated at the end of each sequence to account for drifting of the field induced by the Earth's rotation.
Consequently, over the course of a night, a strip of fixed declination would be scanned for RSOs, resulting in wide-ranging (albeit sparse) coverage of the visible GSO region.

The reduced dead time afforded by the shorter readout and slew times of the RASA allowed for more exposures to be acquired during each pointing, such that for every sequence of seven INT exposures, the RASA was able to acquire up to 16 images in the equivalent period.
Furthermore, the RASA's larger FOV meant that multiple INT pointings could be contained within a single RASA pointing, giving GSO candidates the opportunity to remain in view for longer.
The relative merits of the two instruments for surveying the GSO region are discussed further in the sections that follow.

\section{Astrometric calibration}
\label{sec:astrometric-calibration}

For the survey, custom pipelines were developed to analyse the INT~\citep{blake2019optical} and RASA~\citep{blake2020supplementing} datasets, performing the tasks of image reduction, star trail extraction, astrometric and photometric calibration, and RSO detection and characterisation.
The pipelines are written in Python 3 and employ a variety of existing scientific packages, including: 
\texttt{Astrometry.net}~\citep{lang2010astrometry}; \texttt{astropy}~\citep{astropy2022astropy}; \texttt{astroquery}~\citep{ginsburg2019astroquery}; \texttt{DONUTS}~\citep{mccormac2013donuts}; \texttt{NumPy}~\citep{harris2020array}; \texttt{photutils}~\citep{bradley2016photutils}; \texttt{SciPy}~\citep{virtanen2020scipy}; and \texttt{SEP}~\citep{barbary2016sep}.

\begin{figure}[tbp]
\centering
\includegraphics[width=\textwidth]{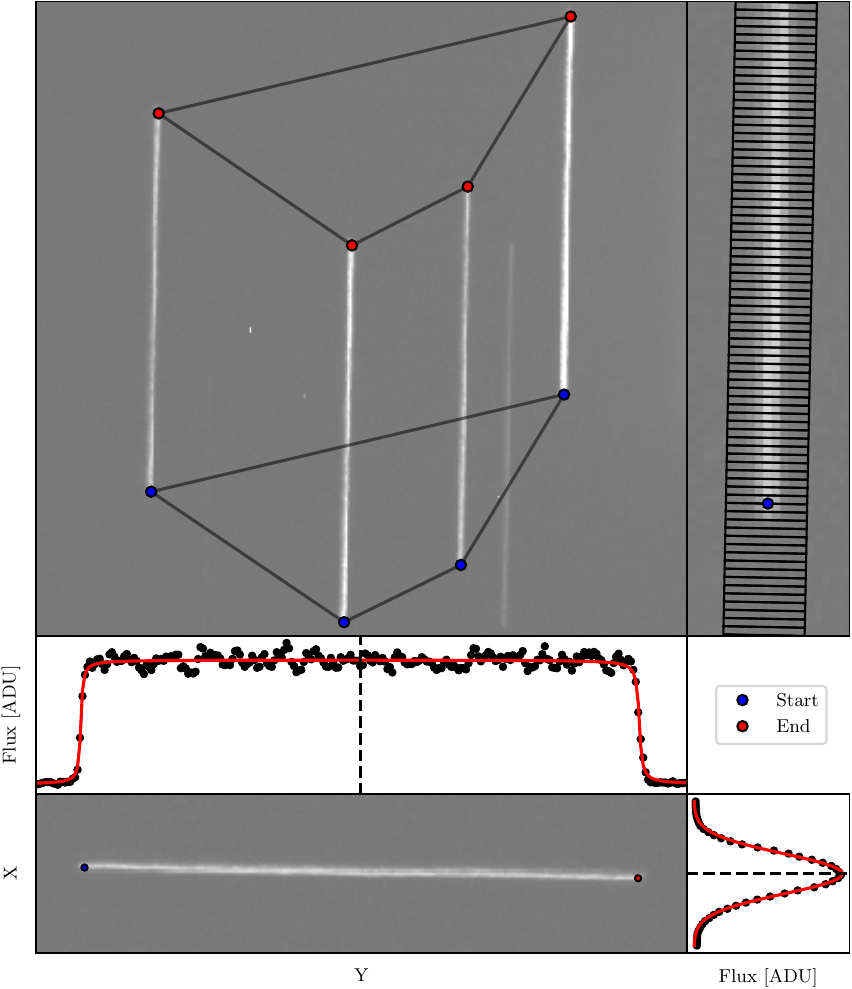}
\caption{Key elements of the updated astrometric calibration routine for the INT frames.
(Top left) Example subset quadrilateral of star trails, labelled with extracted start (blue) and end (red) positions.
(Top right) Zoomed-in panel showing the start point of the left-most trail, with pixel-wide rectangular apertures overlaid (black).
(Bottom) Zoomed-in panel showing the left-most trail rotated by 90$^\circ$, alongside the corresponding cross- and along-trail profile fits.}
\label{fig:astrometry-star-trail-fitting}
\end{figure}

Here, we seek to improve the astrometric calibration for the INT dataset, which proved to be particularly challenging for three key reasons.
Firstly, the star trails are much longer for the INT frames, owing to the smaller pixel scale of the instrument.
With the telescopes fixed and the stellar background accordingly streaking through the images at the sidereal rate, the 10\,s survey exposures would result in star trails of roughly 228 pixels in length for the INT, compared to approximately 96 pixels for the RASA.
The longer star trails amplified the uncertainties associated with the trail centroiding, predominantly affecting the along-trail (right ascension) component.
Furthermore, the INT's small FOV meant that relatively few star trails were detectable on each chip of the CCD mosaic, with typically only $10-20$ trails per chip surviving quality control filtering and checks for blending with other stars.
To combat this issue and minimise the impact of poorly determined centroids (e.g., due to blending or variability), the astrometric fitting was performed in mosaic (full detector) coordinates, as opposed to individual chip coordinates, introducing additional uncertainty associated with the coordinate transformation.
Lastly, the greater extent of the INT star trails served to accentuate the effects of non-linear `pincushion' distortion away from the optical axis of the focal plane, introduced by the telescope optics, alongside wind shake, when centroiding and determining the start/end points of the trails.

The original astrometric calibration routine developed to obtain world coordinate system (WCS) solutions for the INT frames paired \texttt{SEP}, a Python library for source extraction, with the plate solving software, \texttt{Astrometry.net}.
The \texttt{SEP} library is derived from \texttt{SExtractor}~\cite{bertin1996sextractor}, which was primarily developed to extract sources in large-scale galaxy survey images.
It is consequently most proficient when presented with sidereally tracked data, extracting/de-blending point-like and slightly extended (e.g., elliptical) sources.
With the long star trails that result from the observational strategy discussed in Section~\ref{sec:instrumentation-strategy}, \texttt{SEP} can often mistakenly extract multiple sources along a trail or return an inaccurate trail centroid, especially in cases where wind shake has caused the trail to kink or exhibit misleading brightness variability.
We address the first issue by disabling \texttt{SEP}'s object de-blending functionality and applying a filter to retain only those extracted sources that match (within some margin of error) the expected star trail morphology: the orientation of the trail should be aligned with the right ascension (longitudinal) axis; the length of the trail should be close to its expected length based on the exposure time, detector plate scale, and sidereal rate.

\begin{figure}[tbp]
    \centering
    \includegraphics[width=\textwidth]{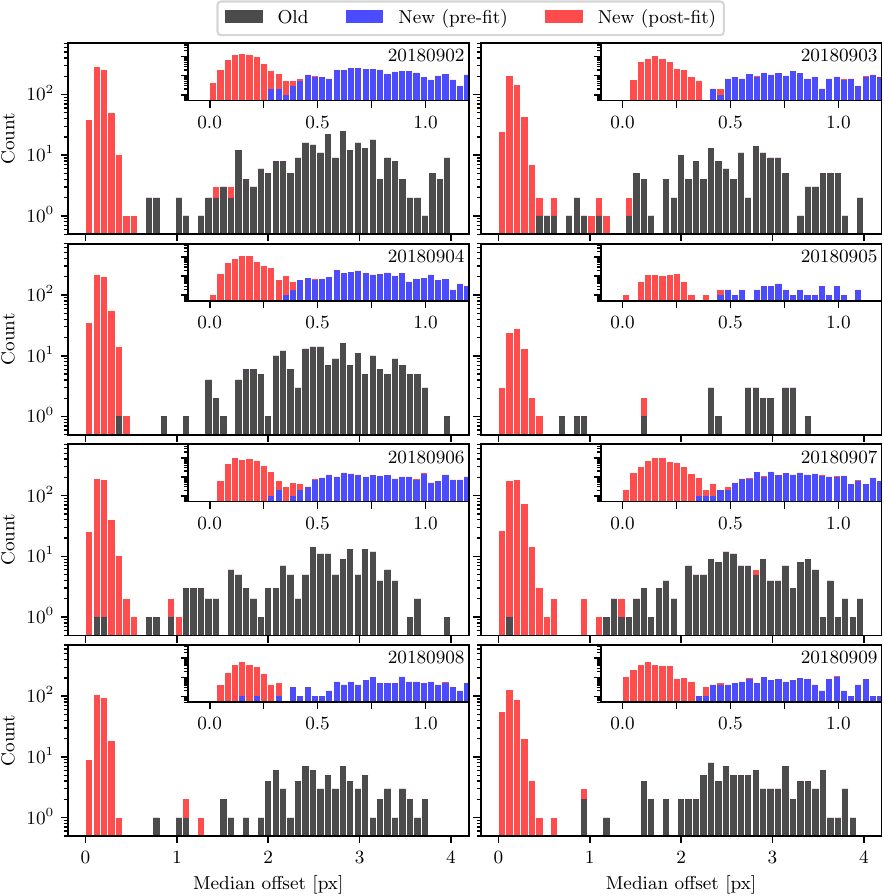}
    \caption{A comparison of the old (black) and new (red) astrometric calibration pipelines for the INT frames.
    We plot the median offsets in right ascension--declination space between the extracted star trail positions and their matched comparison stars, for each night of the survey.
    Inset within each panel, we also provide the corresponding offsets for the pre- (blue) and post-fitting (red, equivalent to `new' above) stages of the new pipeline.
    Note that there are considerably fewer survey frames for the night 20180905, due to poor weather conditions.}
    \label{fig:astrometry-int-nights}
\end{figure}

With the surviving sources, we subsequently carry out a centroid refinement procedure that was previously developed for the shorter trails of GSO detections~\cite{blake2021debriswatch}, fitting the along-trail and cross-trail profiles to obtain more accurate estimates of the trail centroid and its start/end points, as demonstrated in Fig.~\ref{fig:astrometry-star-trail-fitting}.
We fit the cross-trail profile with a Gaussian and the along-trail profile with a `Tepui' function, defined as the difference between two arctangents~\cite[see, e.g.,][]{montojo2011astrometric},
\begin{equation}
    I(x)=A\left[\arctan(b_1(x-c-x_0))-\arctan(b_2(x+c-x_0))\right],
\end{equation}
where $A$ is the normalised amplitude, $b_1$ and $b_2$ are related to the profile tilt, $c$ gives the half-width and $x_0$ is a translational offset. 
We discard trails where the refined centroid differs substantially from the \texttt{SEP} centroid as likely blends.
For the remaining star trails, we use the refined centroids to place rectangular apertures with \texttt{photutils}, obtaining an estimate of flux for the photometric calibration stage of the pipeline.

The refined star trail positions are fed as input to a local installation of \texttt{Astrometry.net} to generate an initial WCS solution, employing the 5200 series of index files, built from the Tycho-2~\cite{wright2003tycho} and \textit{Gaia} DR2~\cite{brown2018gaia} catalogues.
Separately, we perform our own cross-matching of the star trail positions with the \textit{Gaia} DR3 catalogue~\cite{vallenari2023gaia}, making use of the k-d tree matching routines implemented by \texttt{astropy}~\cite{price2022astropy}.
The \textit{Gaia} catalogue represents the current `gold standard' for astrometry, with median positional uncertainties less than 0.1 milliarcseconds (mas), median proper motion uncertainties less than 0.1 mas/yr, and median photometric uncertainties less than a few mmag for the stars used as cross-matching candidates (\textit{Gaia} $G<17$).
We discard comparison stars lacking proper motion information, along with those blended with other stars, based on a consideration of the expected star trail morphology.

\begin{figure}[tbp]
    \centering
    \includegraphics[width=\textwidth]{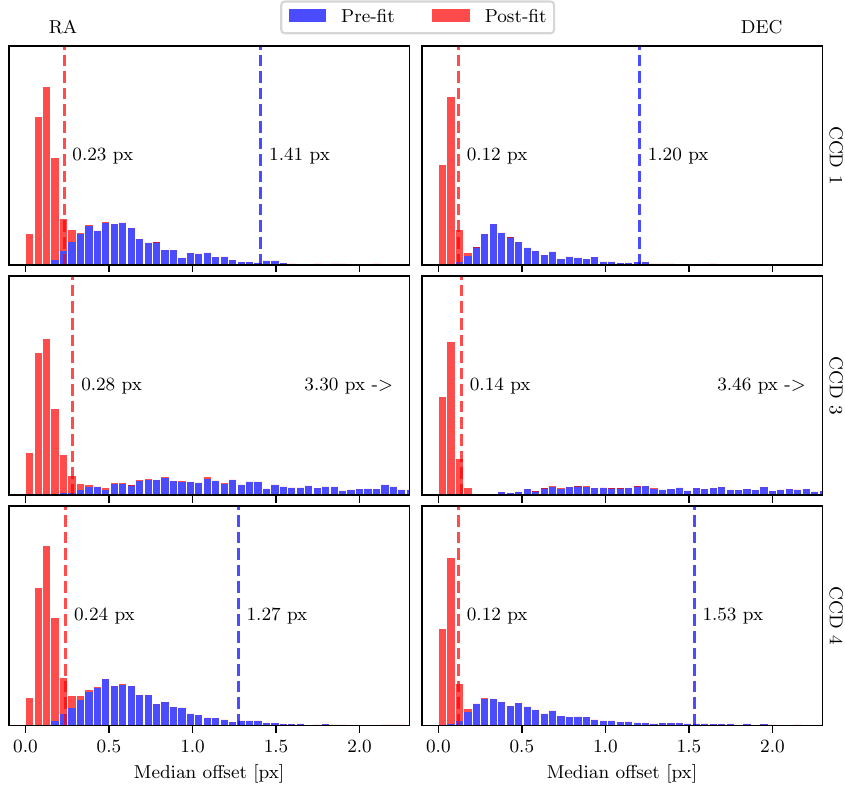}
    \caption{Pre- (blue) and post-fit (red) measures of the astrometric accuracy for the INT survey frames.
    We plot the median offsets in right ascension (top) and declination (bottom) between the extracted star trail positions and their matched comparison stars, for each usable CCD chip.
    Dashed lines indicate the 95$^\text{th}$ percentiles for each case.
    Note that the pre-fit quality for CCD 3 is comparatively very poor due to strong pincushion-like distortion.}
    \label{fig:astrometry-int-chips}
\end{figure}

Previously, we made use of the American Association of Variable Star Observers (AAVSO) All-Sky Photometric Survey (APASS) catalogue~\cite{henden2016vizier}, to closely match the response of the Harris $V$ filter used for the INT observations.
To instead match the $V$-band measurements with the $Gaia$ photometric passbands ($G$, $G_\text{BP}$, $G_\text{RP}$), we make use of the following photometric relationship between \textit{Gaia G} and Johnson $V$, derived by Carrasco \& Bellazzini~\cite{carrasco2023photometric},
\begin{equation}
    y=-0.02704+0.01424x-0.2156x^2+0.01426x^3,
\end{equation}
where $y=G-V$ and $x=G_\text{BP}-G_\text{RP}$.
The resulting zero point distribution, obtained by taking the difference between the catalogue magnitude and the instrumental magnitude for each star trail, is sigma-clipped, and trails with a poor match, either spatially or photometrically, are discarded.
The surviving matches are subsequently used to fit for pincushion distortion, particularly prevalent in CCD 3, positioned furthest from the optical axis of the instrument.
We then perform an iterative improvement of the crossmatch to account for remnant outliers, using the algorithm developed by Chote et al.~\cite{chote2019precision} and applied to the RASA dataset by Blake et al.~\cite{blake2020supplementing} (see Appendix~\ref{app:rasa-astrometry}), obtaining refined astrometric solutions for the start, middle, and end of exposure.

In Fig.~\ref{fig:astrometry-int-nights}, we compare the new astrometric calibration pipeline to its predecessor. 
The new routines achieve sub-pixel accuracy across all eight nights of the survey, in every case a vast improvement on the original astrometry.
Typically, we see a significant reduction in the median separation between detected star trails and matched comparisons, from around 2$''$ to a small fraction of an arcsecond.
The negative impact of the distortion on the quality of the pre-fitting astrometric solution for CCD 3 is clearly evidenced in Fig.~\ref{fig:astrometry-int-chips}.
Nevertheless, we find that the distortion fitting succeeds in bringing the astrometric accuracy back in line with that of the other two CCD chips comprising the usable portion of the WFC mosaic.
For the vast majority of frames, the new astrometric calibration pipeline achieves a median offset (between the extracted star trail positions and their matched comparison stars) below 0.3\,px, equivalent to roughly 0.2$''$.
As expected, the post-fit accuracy is higher in the cross-trail (declination) direction than in the along-trail (right ascension) direction, owing to the greater positional uncertainty introduced when fitting an elongated profile instead of that of a point source.
While the pre-fit accuracy is similar in both directions, the dominant source of uncertainty in declination is due to the distortion; the component associated with the centroiding routine is comparatively small.

\begin{figure}[tbp]
    \centering
    \includegraphics[width=\textwidth]{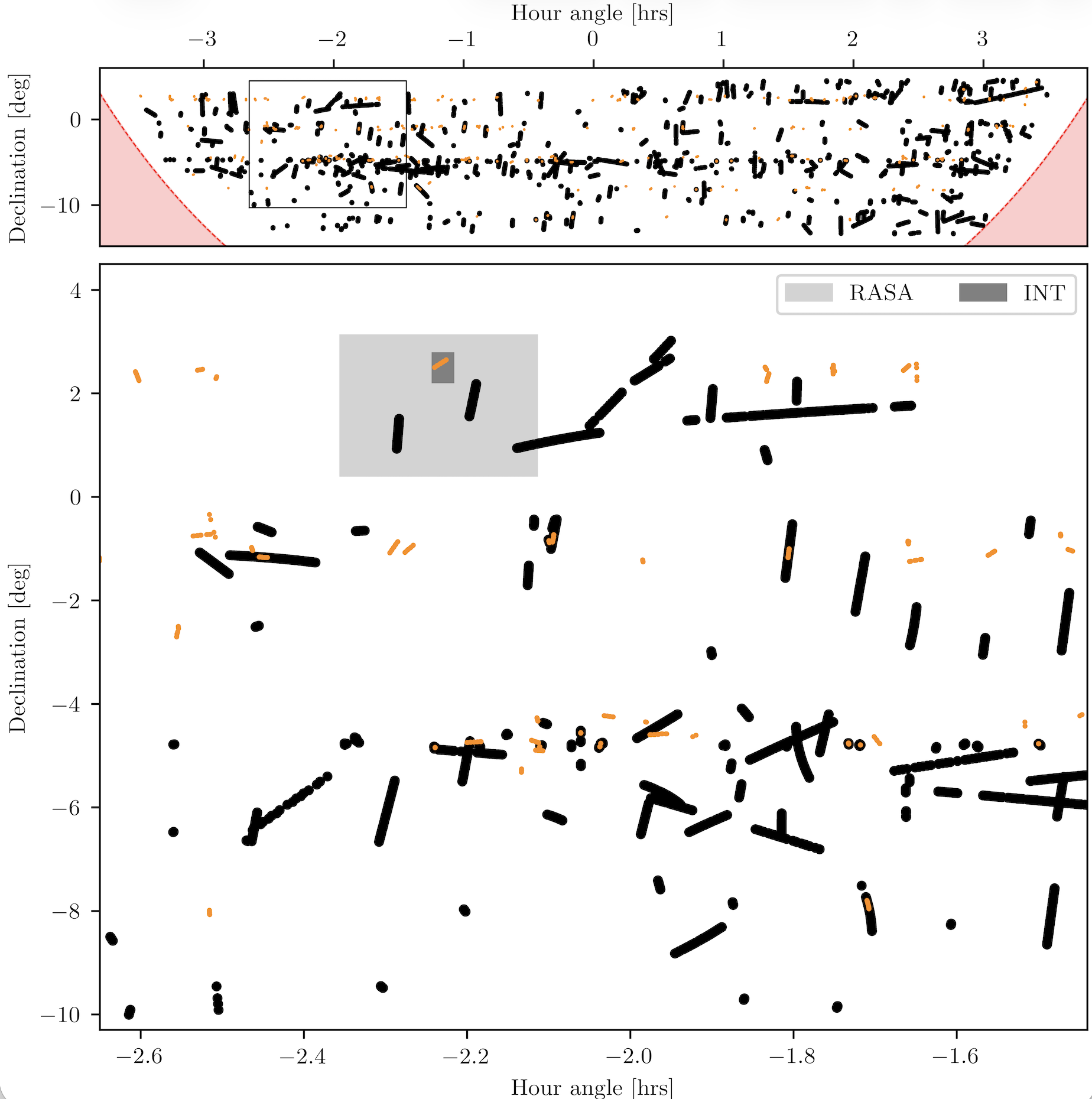}
    \caption{Extracted positions in hour angle--declination space for targets of interest detected in the RASA (black) and INT (orange) survey frames.
    The top panel shows the full sky coverage (all detections) for the survey, illustrating the declination strip scanning strategy discussed in Section~\ref{sec:instrumentation-strategy}.
    The red shaded regions indicate the 30$^\circ$ elevation limit constraining the INT pointing.
    A zoomed-in view of the boxed region is provided in the bottom panel.
    There, we highlight the relative sizes of the INT (dark grey) and RASA (light grey) fields of view.}
    \label{fig:int-rasa-arcs}
\end{figure}

With the new solutions in place, we update the astrometry and photometry for the original INT detections (see Appendix~\ref{app:updated-int-photometry}). 
The detected arcs from both the INT and RASA datasets are shown in Fig.~\ref{fig:int-rasa-arcs}.
Here, we can see how the differing properties of the instruments combine with the chosen observational strategy to yield, in many ways, complementary results.
The RASA's FOV was sufficiently wide to contain multiple pointings of the INT, often extending the coverage of targets to more than $10-15$\,minutes, enabling much longer arcs to be extracted for the RASA detections.
The wider FOV also enhanced the RASA's overall sky coverage throughout the survey, with many bright RSOs being detected outside the on-sky footprint of the INT.
Also evident are a significant number of isolated INT arcs that were too faint to be recovered by the smaller aperture RASA.
In the following section, we apply a blind stacking technique to boost the recovery of faint RSOs in both datasets.

\section{Blind stacking}
\label{sec:blind-stacking}

To recover additional targets in the INT and RASA datasets, we employ an adapted version of the blind stacking algorithm developed by Cooke et al.~\cite{cooke2023simulated}. 
The methodology is designed to extract moving targets from sequences of observed frames and, crucially, is sensitive to targets with motion spanning a range of potential orbits within a single dataset. 
Blind stacking has been shown to recover fainter targets than single-frame object detection techniques; thus, we use it to improve the sensitivity limits achieved by the original pipeline.
We summarise the procedure here, referring the interested reader to the cited paper for a more in-depth discussion.

\subsection{Technique}
\label{sec:blind-stacking-technique}

The blind stacking algorithm takes in a set of pre-processed data frames (the frames have been reduced, and any significantly noisy regions have been masked) and combines them into a single array of $n \times N_x \times N_y$ pixels, where $n$ is the number of frames per stack, and $N_x \times N_y$ is the size of an individual frame. 
Frames are binned by a factor $b$ (averaged in $b \times b$ square-pixel bins) and capped before combining, such that pixel values above or below a threshold are clipped. 
We define a minimum and maximum path length to test, corresponding to the minimum and maximum allowed motion of a target between successive frames, measured in pixels, and then carry out the blind stacking for each path in $(x,y)$ space that falls within the specified range.

\begin{figure}[tbp]
\centering
\includegraphics[width=\textwidth]{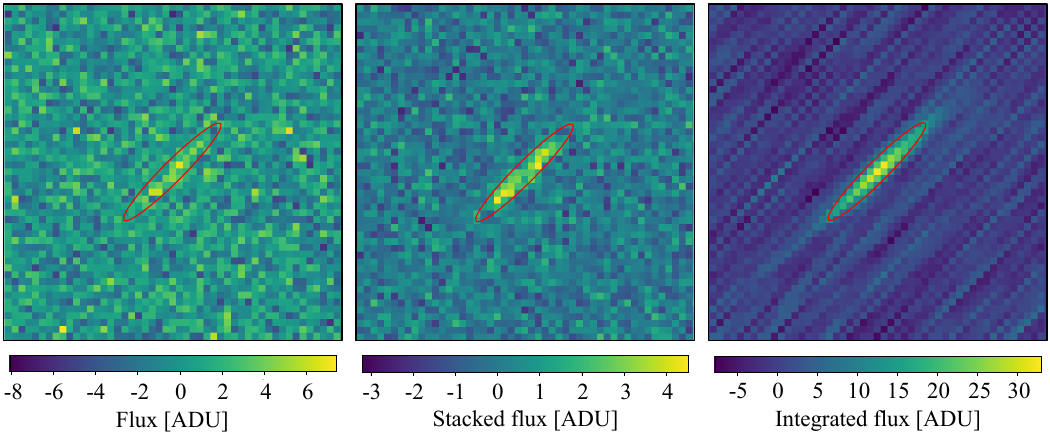}
\caption{The detection of a simulated streaking target across 9 frames.
The simulated target is shown in an individual frame (left), a stack of 9 frames (middle), and an integrated stack of 9 frames (right).
In all panels, the red aperture locates the position of the target.}
\label{fig:example-stacking-simulated}
\end{figure}

For each pixel in the first frame of the stack, we take the selected path and determine the position of the corresponding pixel in each of the other frames. 
We select these $n$ pixels and combine them into a single average value. 
The average can be computed in a variety of ways; in the original algorithm, we employed a clipped mean. 
After repeating this process for every pixel in the frame, we perform a secondary integration step. 
For each pixel, we assume a streak length from the chosen path and integrate all pixels along this theoretical streak, attempting to integrate the flux of any detections. 
We then compare the results of testing every path on a pixel-by-pixel basis, keeping only the brightest pixel at each location alongside the path that produced it. 
The result is a single frame of $N_x \times N_y$ pixels, with brightness spikes indicating the presence of targets moving between successive frames. 
Finally, we search the master frame for detections using \texttt{SEP}, extracting their positions and motions.
In Fig.~\ref{fig:example-stacking-simulated}, we illustrate the described steps using a simulated target with regular motion across 9 frames. 
The frame backgrounds are generated using values drawn from a zero-centred normal distribution to simulate the noise, and the target is a Gaussian streak. 
The left-hand panel of Fig.~\ref{fig:example-stacking-simulated} shows the target in an individual frame, while the middle panel demonstrates a significant improvement in the target's SNR from 0.29 to 0.84, achieved by median stacking 9 frames along the correct path. 
The right-hand panel of Fig.~\ref{fig:example-stacking-simulated} shows a further boost in SNR to 2.59, following the secondary integration step.
In practice, the improvement will not be as stark, as many other paths are also tested, but the simulation suffices to demonstrate the methodology employed.

The INT dataset differs from the simulated dataset used by Cooke et al.~\cite{cooke2023simulated} in a number of ways; therefore, it proved necessary to make some updates and adaptations to the code.
First, for each exposure, we combine data from the four-CCD mosaic into a single frame, allowing the blind stacking algorithm to track an RSO that traverses multiple chips during a telescope pointing. 
We bin the combined frames by a factor of 2, based on a consideration of the instrument's point spread function and the computation times required (computation time scales linearly with the number of pixels per frame).
The next key difference is the comparatively long exposure time used for the survey.
As discussed in Section~\ref{sec:astrometric-calibration}, this resulted in significant stellar streaking within the individual frames.
To ensure that the stars do not contribute to false positive detections, we mask the streaks using the same mathematical morphology technique employed for the original single-frame approach~\cite{blake2021debriswatch,laas2009new}.
Masked pixels are replaced by zeros, treating them as having neither a positive nor negative influence on the stack.
Thirdly, the number of frames per stack is much lower for the INT dataset, a consequence of the observational strategy employed.
To maintain a constant number of frames per stack and to allow the maximum number of pointings to be considered, we opt for rolling stacks of $n=5$ frames, enabling us to corroborate detections between two or more stacks per pointing to mitigate false positives. 
The small number of frames per stack enabled a switch from a clipped mean to a median stack, typically more robust against individual noisy frames, without increasing the pipeline runtime beyond reasonable limits.
Lastly, the simulated dataset in \cite{cooke2023simulated} benefited from the use of Complementary Metal-Oxide-Semiconductor (CMOS) detectors, achieving negligible gaps between frames, whereas dead time was much higher for the INT dataset due to readout of the CCD mosaic.
The stacking stage of the algorithm tests paths that depend on the distance a target could move between successive frames, making the full cadence (the sum of the exposure and readout times) the relevant time step.
The integration step, on the other hand, depends on the length of the target's streak within an individual frame, such that the relevant time step is solely the exposure time.
Due to this, we must rescale each path to be tested between the stacking and integration steps of the pipeline. 
The scaling factor is given by $\frac{e}{c}$, where $e$ is the average exposure time of the $n$ frames used in a stack and $c$ is the average cadence, both calculated independently for each stack.

Many of the above alterations apply to the RASA dataset as well, though there are some differences.
Firstly, the RASA instrument used a single large-format CCD, so the stitching step was not necessary when processing the RASA data. 
Secondly, the RASA frames are significantly larger than their INT counterparts, requiring increased levels of computation for processing.
Lastly, the cadence and pixel scale for the RASA dataset differ from those of the INT: the cadence is shorter, with frames separated by roughly 4\,s of readout; and the pixel scale is larger.
Thus, a target moving at the same rate in the shared field of the two instruments will cross fewer pixels in the RASA frames compared to those of the INT.
This means that the paths to test during the stacking stage of the pipeline are shorter, partially offsetting the additional compute requirements brought about by the larger frames. 
We are therefore able to keep the full median stack of pixel values, as used for the INT dataset. 
The shorter cadence also results in more frames per pointing of the telescope, allowing more consecutive frames to be stacked for the RASA dataset: we use $n=11$.

After obtaining a preliminary list of blind stack detections, we attempt to identify any false positives.
Through the vetting of individual frames fed as input to the blind stacking algorithm, it was found that several INT frames exhibit low-level horizontal noise features, likely caused by pickup noise from the readout electronics.
The noise feature has very little impact on the single-frame thresholding approach but generates a high number of false positives when applying the blind stacking algorithm. 
Due to this, we flag any detections associated with the affected frames and only accept them if they are sufficiently bright.
No comparable features were found to impact the stacking algorithm when applied to the RASA dataset, though we observe a high false positive rate stemming from the extreme corners of the RASA frames, likely caused by slight miscollimation of the optics (see Appendix~\ref{app:rasa-astrometry}).
We mask the corners of the RASA frames prior to stacking to combat this issue.
The next step is to compare the number of detections per stack of $n$ frames. 
We perform sigma-clipping to identify any stacks that appear to be outliers producing large numbers of detections. 
Spikes in individual stacks are typically caused by an underlying issue with the data obtained for the corresponding pointing, such as poor weather conditions.
With detections from these outlier pointings rejected, we then attempt to associate detections between neighbouring stacks.
Detections that are sufficiently similar in position, motion, and angle are assumed to be from the same target, recovered across multiple stacks.
To boost reliability, we keep only those detections that are found in at least two stacks, as those caused by noise or other artefacts are unlikely to produce signals with the same rate and direction of motion in multiple stacks.

For each surviving detection, we use the extracted motion to examine the relevant regions of the individual frames associated with the corresponding stack.
Through visual inspection, we identify any remaining false positives that result from contaminants in single frames, such as the long streaks of bright, fast-moving satellites in medium or low Earth orbit.
Notably, the RASA dataset produces far fewer false positives than the INT data.
This is most likely due to the higher value of $n$ used.
Median stacking more frames reduces the impact of noisy signals aligning and producing false positives.
An additional source of false positives is caused by the remnants of imperfectly masked stellar streaks. 
Due to the increased depth available via the blind stacking method, these faint signals can be similar in magnitude to the targets of interest. 
To mitigate these, we paid close attention to the recovered signals that aligned with the motion of the stellar streaks within individual frames and rejected those we deemed to be the result of stellar contamination. 
This feature becomes more of an issue when testing longer path lengths, those that approach the motion of a star between successive frames.

\begin{figure}[tbp]
\centering
\includegraphics[width=\textwidth]{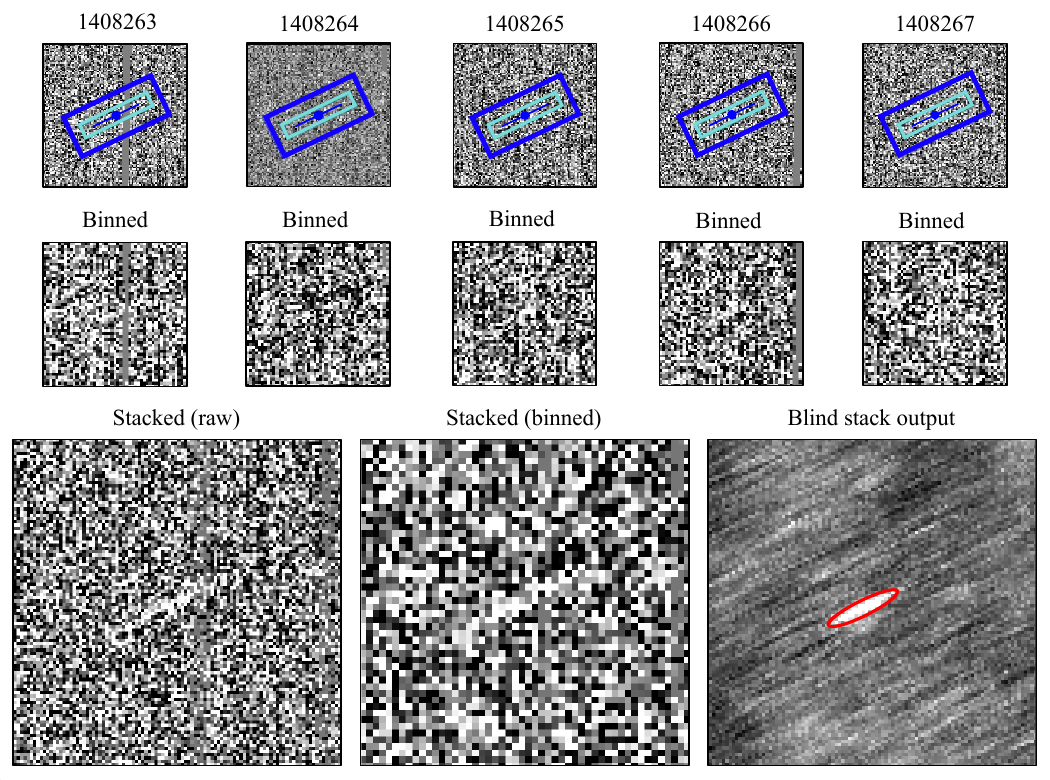}
\caption{An example recovery of a new INT target, with range-corrected brightness $V\sim21$, found using the adapted blind stacking pipeline.
The top row shows five survey frames, while the middle row gives their ($2\times2$) binned counterparts.
From left to right on the bottom row, we show the median stacks of the raw and binned survey frames, alongside the output of the stacking pipeline.
Blue boxes highlight the target location in the individual frames with the blue point and line showing the predicted extent of the detection and the cyan box showing the centroid corrected rectangular aperture. The red ellipse shows the \texttt{SEP} detection in the stacking output frame.}
\label{fig:example-stacking-1408263}
\end{figure}

Finally, we extract photometry for the resulting list of detections.
For each target, we have its position in the first frame of the stack in which it was detected, alongside its motion between frames, and we use this information to locate the target in each frame comprising the stack.
Scaling the inter-frame motion to an intra-frame motion (i.e., by a factor $\frac{e}{c}$) gives the extent of the motion within an individual frame.
We then place a rectangular aperture, with a length determined by the intra-frame motion and a width governed by the point spread function, at the expected position in each frame.
If more than 10\% of the aperture pixels are masked (due to stellar streaks or other features), the flux extracted for that frame is rejected. 
To obtain a single flux estimate for each target, we take the median extracted flux across all accepted frames of the stack.
The photometry extraction stage of the blind stacking pipeline is prone to some uncertainties and imprecision.
Firstly, the target motion recovery is only capable of achieving a resolution on the scale of one pixel within the constituent frames, a consequence of the blind stacking algorithm being forced to offset frames by an integer number of pixels in the x and y directions.
The result is a possible positional error of multiple pixels, especially for the latter frames of the stack.
Additionally, the stacking algorithm assumes a constant exposure time and cadence for each frame within a stack.
For the INT and RASA datasets, this assumption holds for the majority of frames, although some stacks have a distribution of cadences, leading to further uncertainty in aperture placements.
Beyond these algorithm-level concerns, general noise in the data, caused by anything from poorly masked stellar streaks, wind shake, or variability in the target's brightness (see Section~\ref{sec:brightness-variability}), can introduce uncertainty in the recovered position and motion of detections, leading to errors in aperture placement and flux extraction.

\subsection{Detections}
\label{sec:blind-stacking-detections}

In Fig.~\ref{fig:example-stacking-1408263}, we give an overview of the steps described in Section~\ref{sec:blind-stacking-technique} for a new 21$^\text{st}$ magnitude detection recovered from the INT survey frames using the blind stacking technique.
Further examples are provided in Appendix~\ref{app:blind-stacking-examples}, alongside a plot of the flux ranges extracted for each detection.
The target's inferred position is mapped across five individual frames, both raw and binned by a factor of two. 
In the bottom panels of Fig.~\ref{fig:example-stacking-1408263}, we show the corresponding median stacks of the frames and the output of the blind stacking pipeline. 
The boost in SNR is clearly evident in the stacking output frame, and standard threshold-based techniques are subsequently capable of centroiding the target's signal effectively, verifying its motion and enabling the placement of apertures for the extraction of photometry in the individual frames.

\begin{figure}[tbp]
\centering
\includegraphics[width=\textwidth]{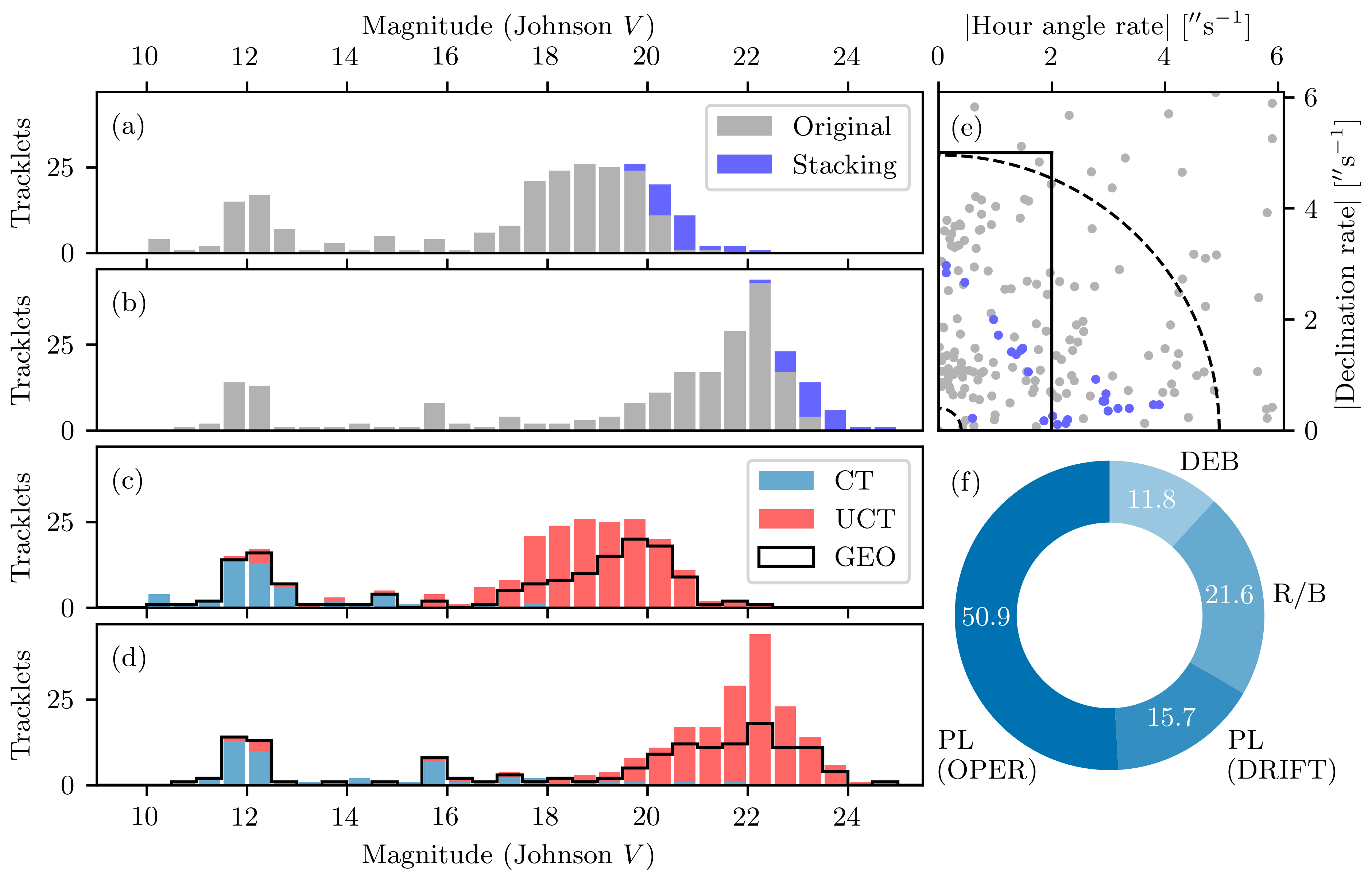}
\caption{(a) Histogram of median calibrated brightness measurements for the full sample of tracklets extracted from the INT survey frames.
Tracklets detected using the original single-frame approach are shown in grey, while those uncovered by the blind stacking technique are highlighted in blue.
In (b), the brightness measurements for the same sample have been normalised by trail length.
Panels (c) and (d) show equivalent histograms to (a) and (b), respectively, though with colours indicating tracklets that correlate with RSOs in the publicly available Space-Track catalogue (CT, cyan) and those that fail to do so (UCT, red).
(e) Angular rates in hour angle and declination for the full INT sample, with blind stack detections highlighted in blue and original detections shown in grey.
Solid black lines indicate a bounding box in rate space, used to extract a circular GSO sub-sample, labelled `GEO' in histograms (c) and (d).
Dashed black lines specify rate limits constraining the blind stacking search.
(f) RSO types for the correlated detections, categorised as operational or drifting payloads (PL), rocket bodies (R/B) or debris (DEB).}
\label{fig:int-brightness-histograms}
\end{figure}

Across the full INT dataset, we recover 25 new detections by applying the adapted blind stacking pipeline.
The stacking detections augment the faint end of the bimodal distribution uncovered by the original analyses~\cite{blake2021debriswatch}, pushing the sensitivity limit of the survey fainter by roughly 1 magnitude, as evidenced in panels (a) and (b) of Fig.~\ref{fig:int-brightness-histograms}.
Unsurprisingly, the new detections fail to correlate with known RSOs in the publicly available Space-Track catalogue~\cite{spacetrack}; the correlation results for the full sample are provided in panels (c) and (d) of Fig.~\ref{fig:int-brightness-histograms}.
As a result, the breakdown of RSO type for the correlated tracklets, provided in panel (f) of Fig.~\ref{fig:int-brightness-histograms}, remains unchanged from the original analyses: over half are associated with operational payloads, many of which were station-kept satellites along the geostationary belt; the remaining tracklets are a mix of decommissioned satellites in graveyard or drifting orbits, abandoned rocket bodies, and debris.
In panels (b) and (d) of Fig.~\ref{fig:int-brightness-histograms}, we have normalised the total flux measured for each detection by a factor $x/l$, where $x$ is characteristic of the point spread function of the optical system, and $l$ is the extent of the angular path traversed by the target during exposure.
The normalisation gives the brightness of a point source with an equivalent surface brightness for the same integration time, thus painting a clearer picture of the detection capability of the sensor.
If we adopt the Optical Size Estimation Model (OSEM) assumptions defined by Africano et al.~\cite{africano2005understanding}, namely a Lambertian spherical phase function, corrected to a phase angle (Sun--target--observer) of 0$^\circ$, and an albedo of 0.1, the stacking algorithm extracts targets as small as $5-10$\,cm in diameter, some of the smallest RSOs observed in the GSO region.
However, these assumptions are highly uncertain, as we lack \textit{a priori} knowledge for detections that fail to correlate with catalogued RSOs.
Moreover, size is one factor among many influencing the photometric behaviour of an RSO, and consequently, the brightness of a detection can vary significantly over the course of the observation window (see Section~\ref{sec:brightness-variability}).

The observational strategy employed for the survey was not designed with stacking in mind; consequently, there are a number of factors that limit the success of the technique when applied to the INT dataset.
The key limitations are the relatively long cadence and the low number of frames per pointing. 
For blind stacking, the number of paths that must be tested scales with the expected motion of a target between successive frames, which similarly scales with the observational cadence.
Every path that is to be tested increases both the computational time and the background level of the master stacked frames. 
Thus, for maximum sensitivity to faint signals, the fewest number of potential paths is preferred.
Furthermore, a significant proportion of the observational cadence comprises dead time between exposures, which has the particularly adverse effect of increasing the number of paths without increasing the quantity of useful data.
Dead time also necessitates the rescaling of paths between the stacking and integration steps of the pipeline, introducing additional uncertainty due to the fact that the cadence is not perfectly uniform across the entire dataset.
The technique also benefits from longer duration pointings of the telescope, as the SNR of a detection increases with the number of frames in a stack (up to a point; see \cite{cooke2023simulated}).
Additionally, with fewer frames per stack, the system is more sensitive to noise features, resulting in a higher number of false positives, as discussed in Section~\ref{sec:blind-stacking-technique}.

Despite the limitations of the dataset, the blind stacking pipeline succeeds in increasing our complement of recovered targets, further probing the faint end of the GSO population.
In panel (e) of Fig.~\ref{fig:int-brightness-histograms}, the angular rates extracted for the new detections are highlighted in blue.
To reduce computational runtime and improve the sensitivity of the method, we place constraints on the motion of prospective targets between frames, thus reducing the number of paths to test.
For the original analyses, we adopted rate limits of $\lvert$Hour angle rate$\rvert<2''$s$^{-1}$ and $\lvert$Declination rate$\rvert<5''$s$^{-1}$ to obtain a sub-sample of tracklets consistent with circular orbits in the GSO regime.
To ensure wide coverage of this region in rate space, we consider motions within the range of $20-270$\,px per cadence, corresponding to targets moving at $0.37-4.97''$s$^{-1}$, as denoted by the dashed lines in panel (e) of Fig.~\ref{fig:int-brightness-histograms}. 
The surviving paths span all orientations, with a resolution of 1\,px (0.66$''$) in both the x and y directions.

We find that the blind stacking performs best for mid-rate targets when applied to the INT dataset, namely those moving at $2-3.5''$s$^{-1}$ through the frames.
The lack of new detections faster than this is unsurprising, as recoverability is known to drop with increasing target velocity~\cite[see, e.g., Fig.~12,][]{cooke2023simulated}, primarily due to reduced signal per pixel as the number of pixels associated with the target rises.
For slow-moving targets, the original pipeline was already proficient in detecting point-like sources and small trails, owing to their superior surface brightness.
With only 5 frames comprising each stack, it is likely that the improvements in SNR are simply not sufficient to supplement the existing yield in the lower region of the rate space in panel (e) of Fig.~\ref{fig:int-brightness-histograms}, especially given the strict vetting procedures in place for false positive removal and the proportionally greater uncertainty introduced by the 1\,px resolution when aligning frames for stacking.
The technique excels for mid-rate cases, where the stacking and integration steps of the blind stacking algorithm are able to extract trails with a surface brightness that previously fell below the sensitivity limit of the single-frame approach.

The RASA observations are better suited to the blind stacking technique for two key reasons: a higher number of frames per pointing of the telescope (and therefore per stack) boosts the SNR of candidate detections, reduces the impact of noisy pixels, and lowers the number of false positives; and a shorter cadence, owing to less dead time between exposures, means that fewer distinct paths must be tested to cover an equivalent region of rate space, resulting in a lower background level in the blind stacking output frame, thus enabling fainter targets to be detected. 
That said, it is important to note that the RASA instrument's larger pixel scale will have negatively impacted the SNR of targets, with the noise floor bolstered by the corresponding increase in sky background per pixel.

\begin{figure}[tbp]
\centering
\includegraphics[width=\textwidth]{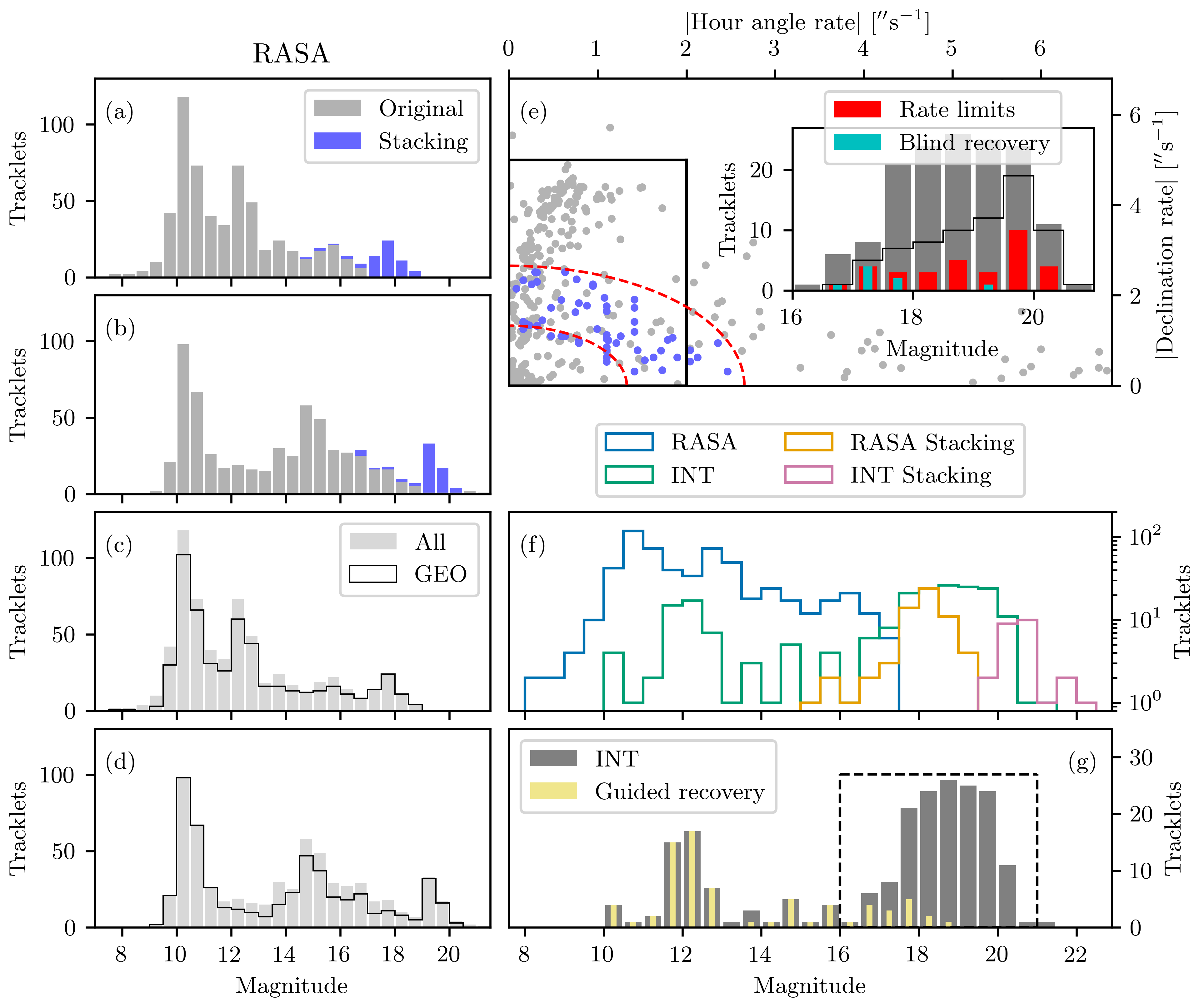}
\caption{(a) Histogram of median calibrated brightness measurements for the full sample of tracklets extracted from the RASA survey frames.
Tracklets detected using the original single-frame approach are shown in grey, while those uncovered by the blind stacking technique are highlighted in blue.
In (b), the brightness measurements for the same sample have been normalised by trail length.
Panels (c) and (d) show the full RASA sample in light grey, overlaid with a solid black step indicating tracklets that satisfy the `GEO' rate limits, defined by the solid black bounding box in panel (e).
There, we plot angular rates in hour angle and declination for the full RASA sample, with blind stack detections highlighted in blue and single-frame detections shown in grey.
Dashed red lines indicate rate limits constraining the blind stacking search.
Inset, we examine the recovery of faint INT detections from the RASA survey frames using the blind stacking technique.
For each brightness bin: the dark grey bar gives the overall number of INT tracklets detected using the single-frame approach; the solid black step indicates the number of those tracklets satisfying the `GEO' rate limits; the red bar specifies the number of tracklets satisfying the blind stacking rate limits; and the cyan bar shows the number of tracklets recovered by the blind stacking algorithm.
The region of interest is highlighted by the dashed box is panel (g) for clarity.
(f) Histograms of median calibrated brightness measurements for the full sample of tracklets extracted from the INT and RASA survey frames, with single-frame yields shown in green and blue, respectively, and blind stacking yields given in pink and orange, respectively.
Note that a colour correction has been applied to the RASA measurements (see main text).
(g) Recovery of INT detections from the RASA survey frames using a guided stacking approach: the grey bars give the overall number of INT tracklets detected using the single-frame approach; and the yellow bars indicate the number of recovered tracklets.}
\label{fig:rasa-brightness-histograms}
\end{figure}

In total, the blind stacking algorithm recovers 87 new detections when applied to the RASA dataset.
Using the extracted motions and fluxes, we associate the detections between consecutive pointings of the telescope, reducing the yield to 62 distinct new RASA detections.
As shown in panels (a) and (b) of  Fig.~\ref{fig:rasa-brightness-histograms}, the stacking detections supplement the faint end of the brightness distribution uncovered by the original single-frame approach~\cite{blake2020supplementing}.
Importantly, the new detections overlap with the ingress of the faint peak in the INT's bimodal brightness distribution; hence, there is an initial rise in the number of tracklets towards fainter magnitudes, followed by a secondary drop-off in sensitivity, roughly 2 magnitudes fainter than that of the original distribution.
The RASA sample is, overall, more numerous than that of the INT, owing to its significantly wider sky coverage.
The single-frame pipeline detected 1205 tracklets, predominantly associated with bright RSOs, with the brightness distribution dropping off for two reasons: initially, at $Gaia$ $G\sim13$, due to the brightness `valley' evident in the INT distribution; and subsequently, at $G\sim16.5$, owing to the sensitivity limit of the relatively small-aperture telescope. 
As for the stacking output, we perform an association of tracklets between neighbouring pointings of the telescope, reducing the yield to 575 distinct detections.
To fully automate the processing of the larger RASA dataset, stricter conditions were put in place for false positive removal, favouring shorter trails that are less likely to be blended or partially detected.
Consequently, the majority of tracklets extracted from the RASA frames satisfy the aforementioned rate limits consistent with circular GSO orbits, as evidenced in panels (c) and (d) of Fig.~\ref{fig:rasa-brightness-histograms}.
Contrary to the INT findings, we find that the bright end of the RASA distribution comprises two separate peaks.
The brightest peak, centred on $G\sim10.5$, is largely associated with station-kept (point-like) geostationary satellites, which were bright enough to saturate pixels and bloom in the INT frames, leading to brightness underestimates and the observed disparity between the two datasets. 
The fainter of the two peaks is primarily attributed to bright derelict RSOs, typically in supersynchronous graveyard orbits and often exhibiting considerable brightness variability (see Section~\ref{sec:brightness-variability}).

Angular rates in hour angle and declination for the RASA sample are provided in panel (e) of Fig.~\ref{fig:rasa-brightness-histograms}, with tracklets detected by the blind stacking pipeline highlighted in blue.
To prioritise the search for very faint RSOs, we dedicated a larger share of available computational resources to the INT dataset, placing narrower constraints on the tested paths when stacking the RASA survey frames.
Based on a consideration of the pipeline performance when applied to the INT dataset, we opt to explore the mid-rate region of parameter space, with a range of motion $12-24$\,px per cadence, corresponding to rates of $1.33-2.65''$s$^{-1}$.
As before, the paths are tested in all directions, with a resolution of 1\,px (1.57$''$) in both the x and y directions.
The superior suitability of the RASA dataset for the blind stacking approach is clear in panel (e) of Fig.~\ref{fig:rasa-brightness-histograms}, with more even coverage of the available rate space and numerous examples of slower-moving targets made detectable by the greater SNR-boost afforded by the higher number of frames per stack.

In panel (f) of Fig.~\ref{fig:rasa-brightness-histograms}, we collate the brightness histograms for the full sample of tracklets extracted from the RASA and INT frames using the single-frame and blind stacking approaches.
The different band passes of the two instruments introduce a systematic offset of approximately 0.5\,magnitudes (see Section~\ref{sec:brightness-variability}), which we account for here.
The precise offset will vary between individual RSOs, depending on their surface properties, attitude, and illumination geometry~\cite[see, e.g.,][]{airey2025comprehensive}; nevertheless, this approximation suffices for a qualitative comparison of the distributions.
The numerical dominance of the bright satellites and rocket bodies extracted from the wide-field RASA frames is immediately clear.
While the faint RASA detections contributed by the blind stacking pipeline are similar in abundance to the INT detections for comparable brightness bins, it is important to remember that the INT covers approximately 2\% of the RASA field, thus yielding many more detections per unit area; a testament to the instrument's superior sensitivity.

\subsection{Recovery}
\label{sec:stacking-recovery}

The contemporaneous INT-RASA observations provide a unique opportunity to evaluate the performance of the blind stacking technique, as they enable us to assess the recovery of INT detections from images of the same fields captured with COTS equipment.

In panel (e) of Fig.~\ref{fig:rasa-brightness-histograms}, the inset histogram examines the recovery of INT detections within the bounds of the rate limits constraining the blind stacking search discussed in Section~\ref{sec:blind-stacking-detections}.
Note that detections brighter than 16$^\text{th}$ magnitude were consistently recovered by both the single-frame and stacking pipelines; thus, we focus our attention on the faint end of the INT brightness distribution.
The red bars of the histogram indicate the number of INT detections falling within the lower and upper bounds of the blind stacking rate limits, denoted by the dashed red lines, while the cyan bars show the number of those detections that are recovered from the RASA frames.
We see that all feasible recoveries brighter than $V\sim17.5$ are achieved through stacking, dropping off for subsequent fainter bins with the sensitivity of the technique.
Interestingly, the faint outlier with $V>19$ is slowly varying in brightness over timescales longer than the INT observation window. 
The outlier was still in view of the RASA during a subsequent pointing, where its brightness had increased sufficiently to be detected by the blind stacking pipeline.

We move on to explore the guided stacking of RASA frames in panel (g) of Fig.~\ref{fig:rasa-brightness-histograms}, leveraging the extracted motions for the INT detections to predict their positions within the RASA frames and inform a median stack.
Inherent uncertainty stems from the INT trail centroiding, especially for cases where the RSO brightness straddles the noise floor (see Section~\ref{sec:brightness-variability}).
That said, the guided nature of the stacking meant that extraction thresholds and false positive vetting could be eased relative to the blind stacking pipeline.
We find that the guided approach is capable of recovering INT detections with $V>18$, highlighting the potential of the blind stacking technique if rate constraints were not in place.

\section{Short arc orbit determination}
\label{sec:short-arc-orbit-determination}

\begin{figure}[tbp]
\centering
\includegraphics[width=\textwidth]{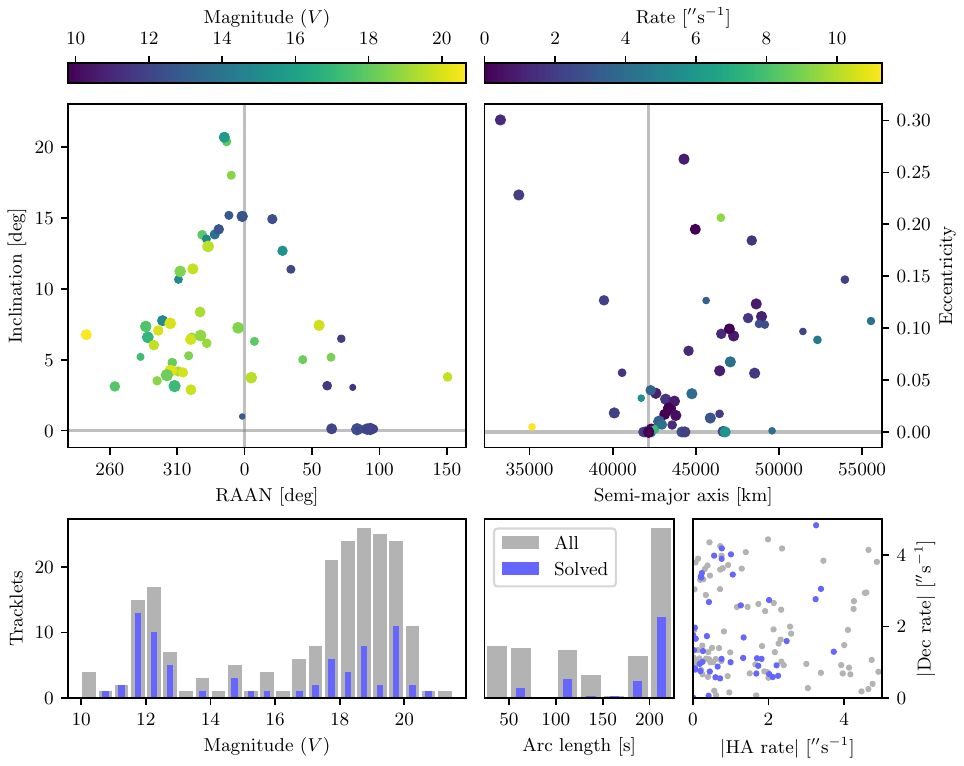}
\caption{Initial orbit determination results for the INT dataset.
(Top left) Right ascension of the ascending node $\Omega$ and inclination $i$ for the solved arcs.
Marker colours indicate the median brightness of the target, as mapped by the colour bar above.
Markers are scaled according to the root mean square of the normalised residuals in right ascension--declination space.
For clarity, grey lines highlight $\Omega=0^\circ$ and equatorial orbits, $i=0^\circ$.
(Top right) Semi-major axis $a$ and eccentricity $e$ for the solved arcs.
Marker colours indicate the rate of the target during the observation window, as mapped by the colour bar above.
Markers are scaled according to arc length.
For clarity, grey lines highlight circular orbits, $e=0$, and the geostationary semi-major axis, $a=42164$\,km.
(Bottom) Histograms of median brightness measurements (left) and temporal lengths (right) for the sample of orbital arcs extracted from the INT survey frames (grey), showing the number of arcs with initial orbit solutions (blue).
Alongside, we plot the absolute rates in hour angle and declination for the sample (grey), with solved arcs highlighted (blue).}
\label{fig:int-orbits}
\end{figure}

The observational strategy detailed in Section~\ref{sec:instrumentation-strategy} was designed to mitigate the narrow FOV of the INT, promote sky coverage, and boost sensitivity for the survey.
In doing so, however, it placed severe limitations on the time available for observing any individual RSO, resulting in very short orbital arcs.

A preliminary initial orbit determination (IOD) assessment was conducted using a least-squares minimisation of normalised right ascension and declination residuals. 
A common initial guess was first adopted, with a semi-major axis equal to 42,000\,km, zero eccentricity, and zero inclination. 
For cases that failed to converge, an alternative initial guess obtained from a surrogate optimisation of the residuals was used. 
The analysis considered observational arcs ranging from roughly 45\,s (two exposures) to 220\,s (seven exposures). 
Only arcs spanning multiple exposures were retained to ensure basic observability. 
Convergence was assessed using a combination of a root mean square error threshold and visual inspection.

The IOD solutions are summarised in the top panels of Fig.~\ref{fig:int-orbits}.
We see that the inclinations $i$ and right ascensions of the ascending node (RAANs) $\Omega$ follow the expected evolutionary trend induced by the perturbative forces acting on RSOs in the GSO region~\cite{schildknecht2007optical,mcknight2013new}.
The Earth's oblateness and its gravitational interactions with the Moon and the Sun cause the orbital planes to precess with a
53-year cycle from $0^\circ-15^\circ-0^\circ$ about the stable Laplacian plane.
The majority of bright detections are found along the zero inclination line, associated with operational geostationary satellites that are N-S station-kept to counter the natural increases in inclination.
On the other hand, the uncontrolled debris have progressed through part (possibly all) of the 53-year cycle, gradually evolving from their assumed origin near the geostationary belt with low inclination and $\Omega\sim100^\circ$, then decreasing in RAAN and increasing in inclination to the maximum $i\sim15^\circ$ at the halfway point of the cycle. 
Many of the faint detections have surpassed this stage, now decreasing in both RAAN and inclination, approaching $i\sim0^\circ$ once more, where the oldest fragments of debris will have already restarted the cycle.
There are tentative signs of clusters in $\Omega-i$ space, typically indicative of groups of fragments with a common explosive origin, though the numbers are too small here to draw any statistically significant conclusions.
We note that, for many of the apparent clusters, the constituent members vary widely in orbital eccentricity, perhaps due to different area-to-mass ratios governing their evolution under solar radiation pressure.

Some degree of spread is to be expected; however, detections straying significantly from the evolutionary path are likely in eccentric orbits.
The IOD solutions span a wide range of semi-major axis and eccentricity values, with arc length playing a key role in solution quality. 
In particular, solutions yielding semi-major axes larger than 50,000\,km are likely to be spurious.
For short observation windows, RSOs in eccentric orbits with their apogees near the GSO region, such as geostationary transfer orbits, can resemble objects in higher circular orbits, thus misleading the solver with degenerate solutions.
These cases are also consistently associated with very short arcs, typically shorter than $\sim2$ minutes, highlighting the intrinsic ambiguity of orbit determination from limited angular data.

The results show a strong dependence of convergence on arc length, with the majority of solved arcs in excess of 200\,s, as illustrated in the bottom panels of Fig.~\ref{fig:int-orbits}.
Using the nominal initial guess, the overall convergence rate is approximately 18\%, increasing to 38\% when the initial condition is instead taken from the surrogate optimisation, with failures in both cases dominated by very short arcs.
We see that the majority of bright tracklets solve; an unsurprising result, given that the bright, point-like nature of geostationary satellites makes them far less likely to drop out due to blending, streaking, or sensitivity.
Interestingly, we observe a slight upward trend in convergence towards the faintest detections, likely due to the fact that the average angular rate of detections decreases as we approach the sensitivity limit of the instrument (a systematic bias stemming from the surface brightness considerations discussed in Section~\ref{sec:blind-stacking}), thus reducing the chance of blending and increasing the likelihood that the targets remain within the FOV for the duration of the seven-exposure pointing.
Overall, while further work using improved data processing (e.g., outlier rejection) and more robust approaches~\cite[see, e.g.,][]{pirovano2020probabilistic} is needed for a comprehensive analysis, this assessment demonstrates the feasibility of IOD and provides a promising basis for future refinement and extension.

\section{Brightness variability}
\label{sec:brightness-variability}

An important finding from the original survey analyses~\cite{blake2021debriswatch} was that many of the faint, uncatalogued RSOs detected by the INT showed significant brightness variability throughout the observation window.
In the following section, we explore this variability more quantitatively.

\begin{figure}[tbp]
\centering
\includegraphics[width=\textwidth]{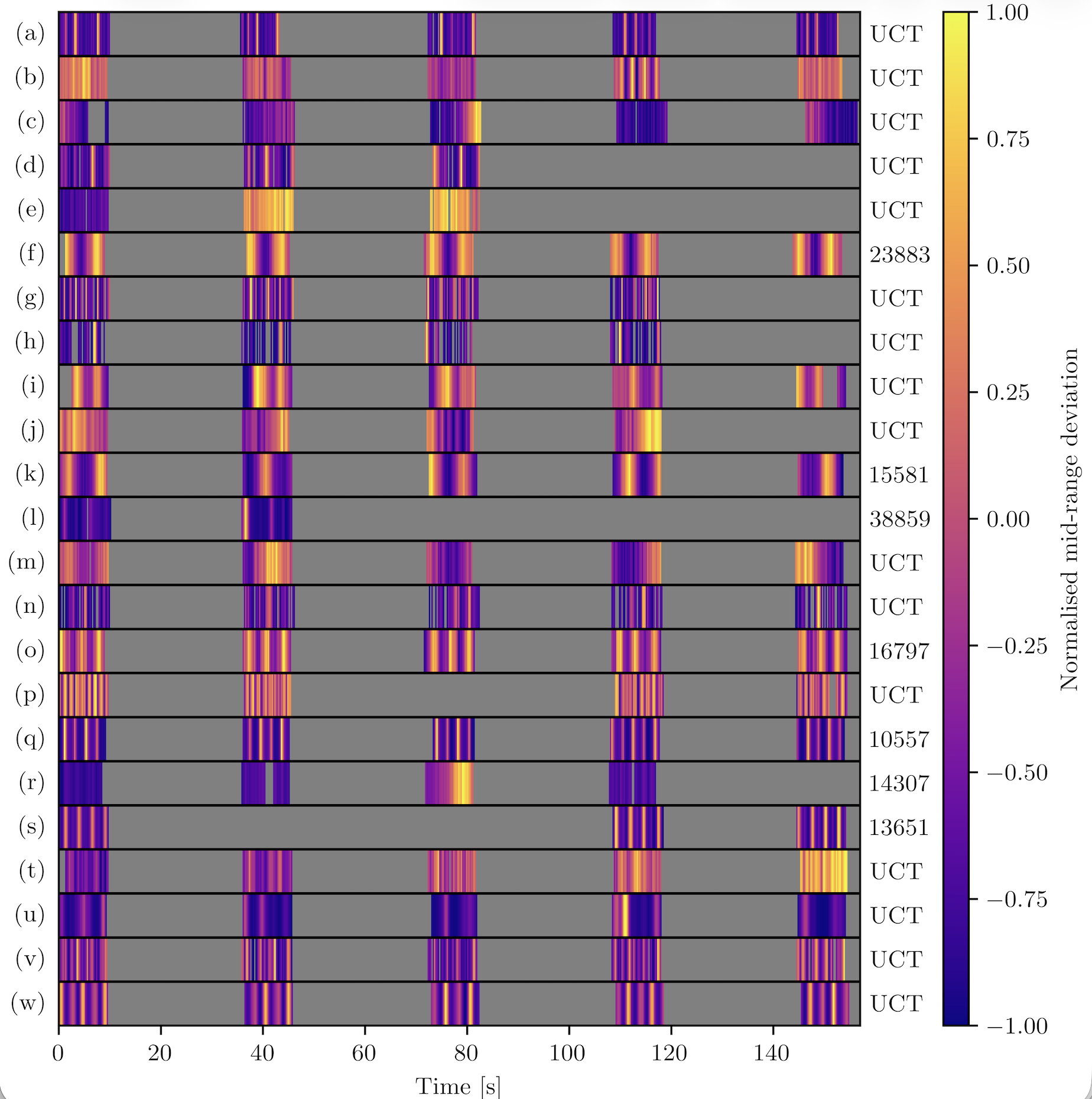}
\caption{Examples of temporal structure exhibited by light curves extracted from the INT survey frames.
The normalised mid-range deviation, $\Delta$ (see main text), is plotted as a function of time relative to the start of the exposure sequence.
Dark colours indicate regions of the light curves that are faint relative to the mid-range, while light colours highlight bright regions.
Grey regions are either inter-exposure gaps due to readout, or intra-exposure gaps arising from blending or bad pixels.
Labels specify the NORAD ID for light curves that correlate with a known RSO in the Space-Track catalogue, or `UCT' for uncorrelated detections.}
\label{fig:int-lightcurves-cmap}
\end{figure}

The reflected light from an orbiting body carries signatures of its shape and attitude, convolved with sensor characteristics, atmospheric effects, and viewing geometry.
Isolating these influences is challenging, and light curve characterisation for space domain awareness remains an active area of research~\cite[see, e.g.,][]{silha2021light}.
Numerous studies have made use of photometric light curves to characterise~\cite{koshkin2024determination,airey2025lord,isoletta2025attitude} and classify~\cite{linares2020space,kerr2021light,shrive2024classifying} objects in near-Earth orbit.

The trailing nature of near-geostationary detections in the INT frames enabled the extraction of high-cadence light curves via the placement of apertures along the trails. 
% Importantly, all faint detections (below V=17) are trailed, owing to the natural perturbative forces at play in the GSO region.
Discrete pixel-wide rectangular apertures were used to avoid correlated noise injection (see~\cite{blake2021debriswatch} for further details).
The light curve extraction method resulted in 10\,s segments (spanning the exposure time) of closely sampled brightness measurements, separated by 25\,s gaps due to the readout of the CCD mosaic, as illustrated in Fig.~\ref{fig:int-lightcurves-cmap}.
To enable a shape-based comparison between light curves that differ widely in mean brightness and variability amplitude, we rescale each brightness measurement by subtracting the mid-range and dividing by the maximum absolute deviation from the mid-range, yielding a dimensionless series $\Delta\left(t\right)\in\left[-1,1\right]$.
Specifically, we plot a normalised mid-range deviation, $\Delta\left(t\right)=\left(f\left(t\right)-M\right)/\Delta_\text{max}$, where $f\left(t\right)$ is the time-dependent, background-subtracted flux extracted via the aperture photometry,  $M=\left(f_\text{max}+f_\text{min}\right)/2$ is the flux mid-range, and $\Delta_\text{max}=\left(f_\text{max}-f_\text{min}\right)/2$ is the maximum deviation from the flux mid-range, where $f_\text{max}$ and $f_\text{min}$ are the maximum and minimum flux, respectively.
Through this interpretation, we can obtain a clearer picture of the variety of variability structures present within the overall dataset while losing information pertaining to the relative amplitudes of variability. 
In Fig.~\ref{fig:int-lightcurves-cmap}, we see evidence of brightness variability on timescales shorter (see, e.g., q) and longer (see, e.g., m) than the exposure time, and indeed longer than the observing window entirely (see, e.g., t), with the nature of the variability ranging from smooth periodic oscillations (see, e.g., f) to sharp sporadic glints (see, e.g., u), often straddling the noise floor of the survey frames.

\begin{figure}[tbp]
\centering
\includegraphics[width=\textwidth]{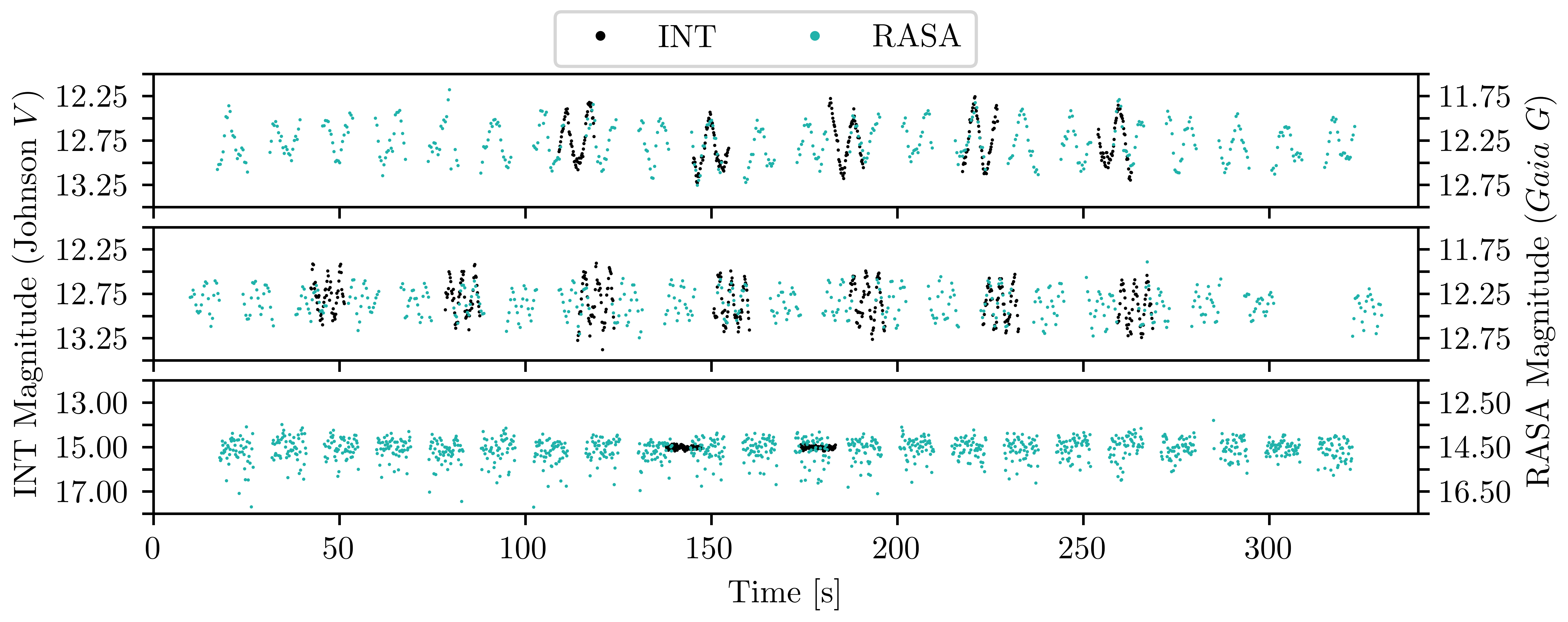}
\caption{Examples of contemporaneous light curves extracted from the INT (black) and RASA (cyan) survey frames.
We apply a 0.5 magnitude correction to align the two sets of measurements, a systematic offset arising from the different bandpasses of the two instruments.
The light curves shown in the top and middle panels are associated with SL-12 rocket bodies NORAD 16797 and NORAD 15581, respectively.
The bottom panel shows the light curve extracted for a tracklet that failed to correlate with the publicly available Space-Track catalogue.
Adapted from \cite{blake2020supplementing}.}
\label{fig:int-rasa-lightcurves}
\end{figure}

Temporal resolution within each light curve segment is governed by the target's angular motion across the CCD chip.
A target moving through the INT field at 1$''$s$^{-1}$ will have traversed roughly 15\,px during the 10\,s exposure, resulting in a 0.66\,s cadence for the corresponding segment.
On the other hand, an angular motion of 5$''$s$^{-1}$ would enable a much higher cadence of 0.13\,s.
This is demonstrated nicely in Fig.~\ref{fig:int-rasa-lightcurves}, where we compare light curves extracted from the INT and RASA frames for RSOs observed contemporaneously.
In general, we find excellent agreement between the extracted INT and RASA photometry after accounting for a systematic offset of $\sim0.5$ magnitudes, attributed to the different bandpasses of the two instruments.
Although the RASA was scheduled to survey the same fields as the INT, the exposure sequences during each pointing were not synchronised.
Consequently, the RASA's superior duty cycle served to fill the large readout-induced gaps in the INT light curves, enabling targets to be monitored over a larger fraction of the observing window.
That said, the light curves also serve to highlight some limitations of the RASA observations.
The temporal resolution is lower for the COTS instrument, owing to its larger pixel scale.
In the top panels of Fig.~\ref{fig:int-rasa-lightcurves}, we see that the RASA's lower temporal resolution acts to suppress the amplitude of variability.
Furthermore, for the fainter detection in the bottom panel of Fig.~\ref{fig:int-rasa-lightcurves}, the brightness measurements extracted from the RASA frames show a high level of scatter, reflecting the lower SNR of sources in the RASA frames, compared to equivalent sources observed by the INT. 
In the following analyses, we consider solely the trailed detections from the INT dataset.
For point-like geostationary targets, it was only possible to extract a single photometric measurement per exposure, yielding much lower-cadence light curves with comparatively limited information over the observing window.

With high-cadence measurements, it is possible to resolve finer details in the target's light curve, such as sharp glints and/or rapid periodicity.
A lower sampling rate acts to suppress this information, smoothing over sharp features and painting a less distinct picture of the underlying variability.
However, an inherent trade-off arises from the light curve extraction method between cadence and signal-to-noise, as reflected light from faster-moving targets will spread across more pixels, reducing surface brightness.
To boost signal-to-noise for such cases, we bin the light curves to sample at a maximum cadence of 0.25\,s.
Care is taken to bin the light curves in linear flux space, before converting to calibrated magnitudes with the updated photometry from Section~\ref{sec:astrometric-calibration}.

\begin{figure}[tbp]
\centering
\includegraphics[width=\textwidth]{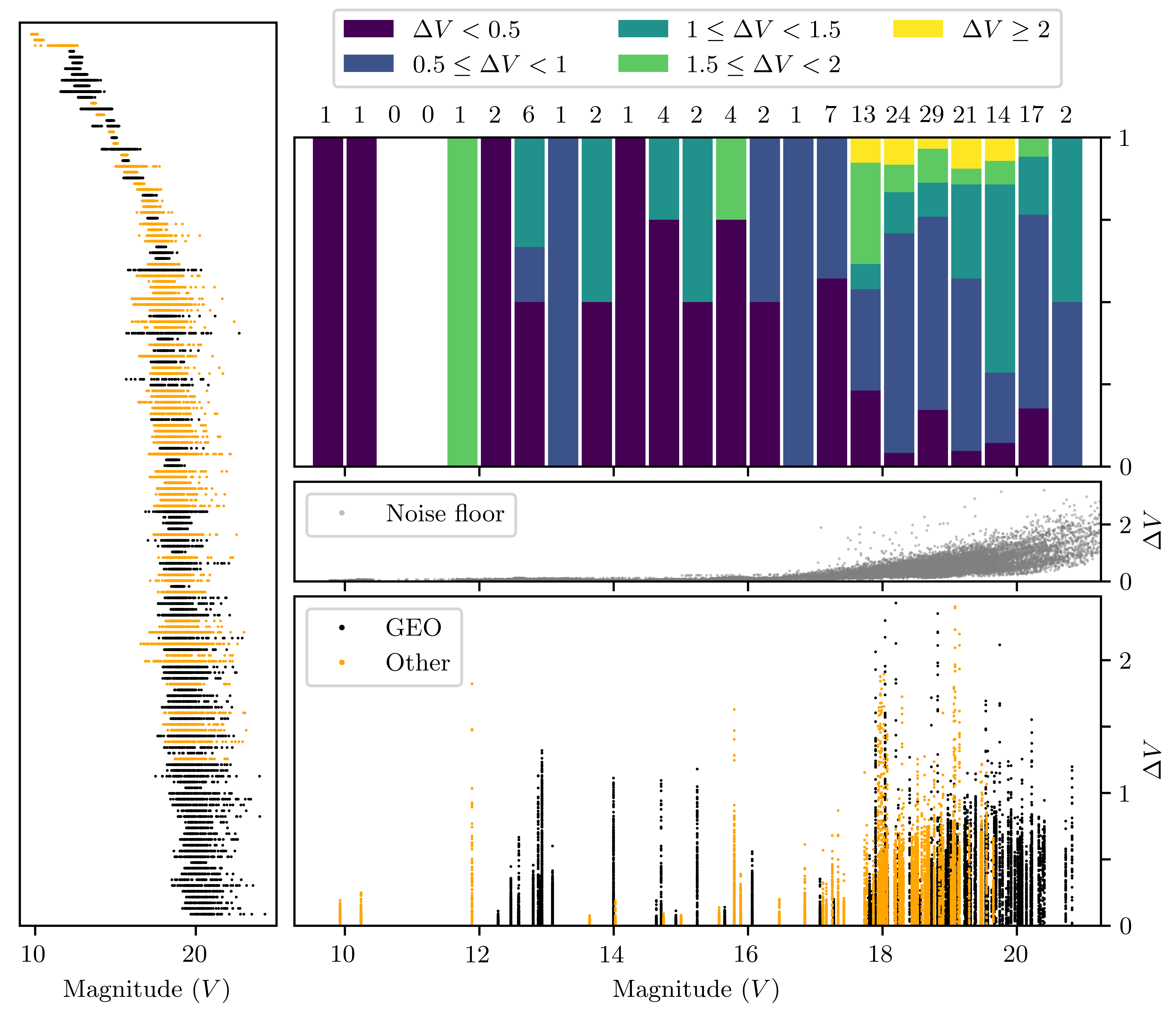}
\caption{Variability exhibited by the photometric light curves extracted for trailed detections in the INT survey frames.
(Left) A ridgeline plot showing the spread of calibrated brightness measurements for the overall sample.
Each row corresponds to a separate light curve, and the rows are sorted by mean calibrated magnitude.
Detections satisfying the `GEO' rate conditions are black, while other detections are shown in orange.
(Bottom right) For each light curve, we plot the difference in brightness $\Delta{V}$ (delta magnitudes) between each brightness measurement and the mean brightness across the observation window, with maximal uncertainty subtracted.
The differences are plotted as a function of the mean brightness for each case, such that each spike represents a separate light curve.
For reference, the photometric noise floor is provided in the middle right panel.
(Top right) The light curves are categorised by the amplitude of their brightness variability, as presented in the bottom right panel.
The number of light curves comprising each 0.5 magnitude bin is labelled above the corresponding bar.}
\label{fig:int-lightcurves-variability}
\end{figure}

In Fig.~\ref{fig:int-lightcurves-variability}, we explore the amplitude of variability for the resampled INT light curves.
The left-hand panel displays a ridgeline plot that shows the spread of calibrated brightness measurements for each light curve, sorted by mean brightness.
We see that the faint end of the sampled population exhibits significant brightness variability, with some examples spanning several magnitudes.
It is also interesting to note that the very faint end ($V>20$) is dominated by slow-moving targets, as expected from the surface brightness considerations discussed above.

Of course, the level of photometric uncertainty increases as we move to fainter signals, and is particularly impactful for targets straddling the noise floor of the observations.
To account for this, we plot the absolute deviation from the mean brightness for the measurements comprising each light curve, with maximal uncertainty subtracted in linear flux space, in the bottom-right panel of Fig.~\ref{fig:int-lightcurves-variability}.
This metric gives a sense of the minimum level of variability that cannot be attributed to photometric uncertainty.
Despite the uncertainty suppression, we find that the faint end remains proportionally more variable than the bright end of the distribution.
Indeed, the normalised histogram in the top-right panel of Fig.~\ref{fig:int-lightcurves-variability} shows that the highest levels of variability are exhibited by detections with a mean brightness in the range $18<V<20$, while the proportion of detections with minor variability drops off considerably towards fainter magnitudes.

\begin{figure}[tbp]
\centering
\includegraphics[width=\textwidth]{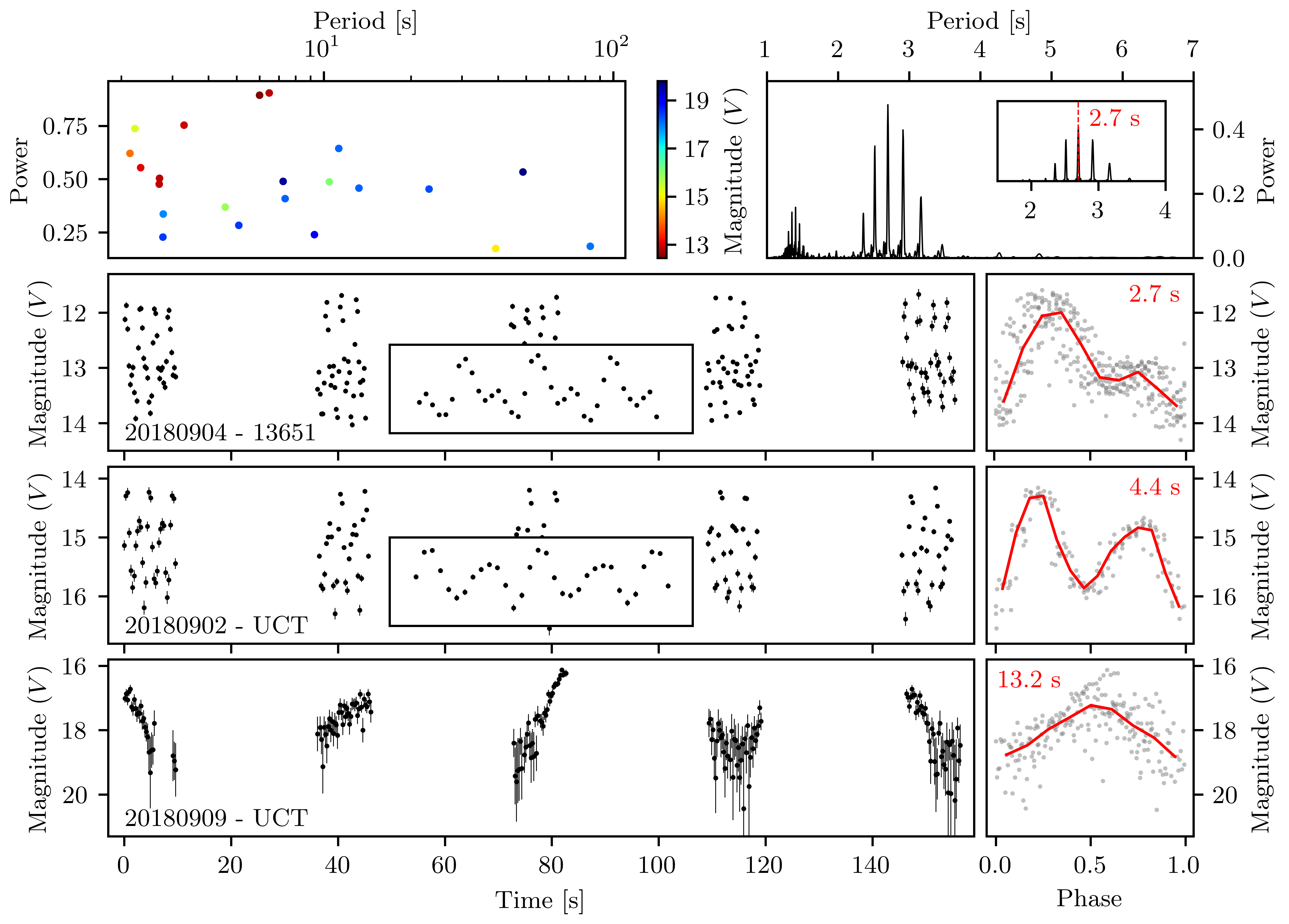}
\caption{Lomb-Scargle (LS) periodogram results for the trailed detections in the INT survey dataset.
(Top left) Peak power and associated period extracted from LS periodograms of the INT light curves, for cases that survive filtering to remove false positives (see main text).
We colour the light curves according to their mean calibrated brightness.
(Top right) Example periodogram for the light curve of a catalogued target, SBS 3 (NORAD 13651).
Inset, we provide the window function to highlight aliasing arising from irregular sampling due to readout-induced gaps in the data.
The bottom three rows give examples of light curves that survive the filtering, the first associated with SBS 3, and the latter two extracted for uncatalogued detections.
Inset, we provide zoomed-in views of 10\,s segments for the top two examples, to more clearly highlight the short-period variability.
Alongside each light curve, we show the associated phase curve, folded using the peak-power period or a more appropriate alias (labelled).}
\label{fig:int-lightcurves-lombscargle}
\end{figure}

We investigate signal periodicity using \texttt{astropy}'s implementation of the Lomb-Scargle (LS) periodogram~\cite{vanderplas2018understanding}.
The LS periodogram remains effective when handling unevenly spaced data with observational gaps, common features of real survey data obtained with ground-based telescopes, as we have seen here.
For each light curve, we extract the peak frequency and corresponding periodicity. 
We then perform a series of checks to filter out false positives.
The extracted frequency must not exceed the Nyquist frequency, and we set a lower limit of $2/T_\text{window}$, where $T_\text{window}$ is the observation window, to avoid spurious signatures of exposure smearing and poorly constrained periodicity, respectively.
For the surviving peaks, we compute the false alarm probability~\cite{baluev2008assessing}, setting a 99\,\% confidence threshold to retain only statistically rare peaks that are unlikely to originate from noise.
We also impose a minimum power threshold using the corresponding 1\,\% false alarm level to ensure that the extracted periodicities are strong relative to the noise floor.
Lastly, we visually inspect the results of phase-folding the surviving light curves using their extracted periodicities, discarding any remaining anomalies.

The top panel of Fig.~\ref{fig:int-lightcurves-lombscargle} shows the periodogram outputs for the light curves that survive filtering.
Approximately half of the light curves with extracted periods less than 10\,s are associated with catalogued (bright), inactive satellites and rocket bodies that are drifting in super-synchronous graveyard orbits.
Defunct GSO satellites commonly exhibit photometric signatures of rapid tumbling, and there is strong evidence to suggest that their spin states evolve predominantly under the influence of the YORP effect~\cite{albuja2018yorp,benson2021radar}.
We also find several examples of faint debris with longer-timescale periodicity, often straddling the noise floor of the observations, as is the case for the example provided in the bottom panel of Fig.~\ref{fig:int-lightcurves-lombscargle}.
Of course, when the brightness drops below the sensitivity limit, we lose access to the target signal (and therefore its variability) until it breaches the noise floor once more.

To test different alias periods, we phase-fold the light curves using the most likely candidates from the periodograms. 
Candidate periods that produce significant scatter in the phase curves are discarded, and the examples provided in Fig.~\ref{fig:int-lightcurves-lombscargle} utilise the best candidate period for each light curve.
Remnant scatter can be explained by a variety of sources: for detections straddling the noise floor, there is significant scatter in the light curves themselves due to higher photometric uncertainty; scatter could result from intrinsic cycle-to-cycle variation, indicative of more complex behaviour; the light curve extraction procedure outlined in \cite{blake2021debriswatch} could result in slight phase offsets between the different segments of the light curve, owing to the uncertainty in time (phase) that stems from the uncertainties associated with the extracted start and end positions for each trail.
The effect of the changing solar phase angle is negligible due to the short total extent of the observing window.

It is important to note that we see many other examples of brightness variability that lack strong periodicity as searched for here, varying from sharp glints piercing sporadically through the noise floor to longer-timescale trends that are not fully captured by the window of observation.
We refer the reader to \cite{blake2021debriswatch} for a diverse montage of photometric signatures extracted from the INT  dataset.

\section{Follow-up observation campaign}
\label{sec:future-work}

The INT survey served as the first instalment of DebrisWatch, a collaboration between the University of Warwick and the Defence Science and Technology Laboratory (UK) investigating the population of GSO debris~\cite{blake2019optical,blake2020supplementing,blake2021debriswatch}. 
Following its success, a second observation campaign took place between March 2022 and January 2023.

\begin{figure}[tbp]
\centering
\includegraphics[width=\textwidth]{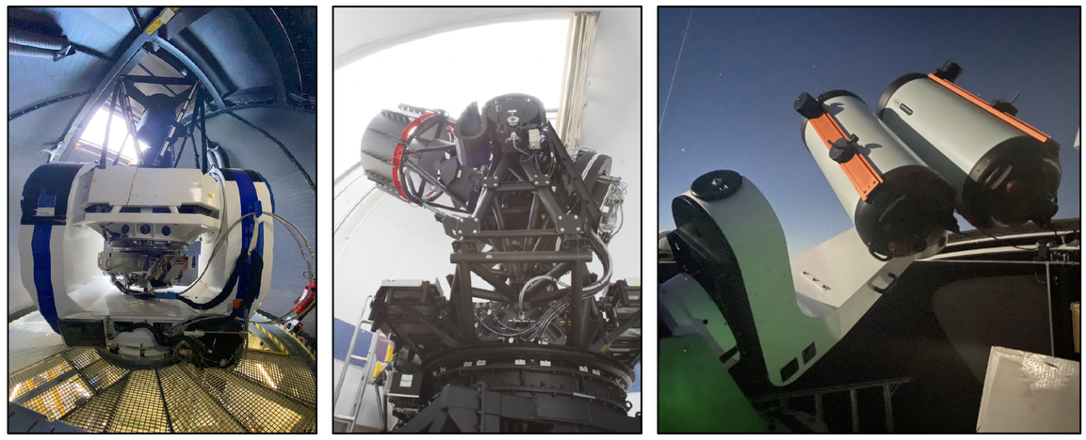}
\caption{Instruments employed for the follow-up observation campaign.
(Left) 1.35\,m SkyMapper Telescope at Siding Spring Observatory, Australia.
(Middle) 1\,m telescope at Bisei Space Guard Center, Japan.
(Right) Twin 36\,cm Warwick CLASP telescope at the Roque de los Muchachos Observatory, La Palma.}
\label{fig:skymapper-bisei-clasp}
\end{figure}

The follow-up campaign placed a greater focus on geographical diversity, utilising time on the following instruments to survey the GSO region over a broader range of longitudes: the 1.35\,m SkyMapper Telescope at Siding Spring Observatory, Australia~\cite{onken2024skymapper}; the 1\,m telescope at Bisei Space Guard Center, Japan~\cite{isobe2004new}; and the twin 36\,cm Warwick CLASP telescope at the Roque de los Muchachos Observatory, La Palma~\cite{cooke2023simulated}.
The three instruments are shown in Fig.~\ref{fig:skymapper-bisei-clasp}, and specifications are provided in Table~\ref{tab:instrument-specification-followup}.
As for the original survey (see Section~\ref{sec:instrumentation-strategy}), the follow-up observations were conducted at low solar phase angles with the telescopes stopped, mapping a strip of constant declination over the course of the night.
While these elements of the observational strategy remained the same, the exposure sequences comprising each pointing of the telescope were modified to better exploit the characteristics of the available instruments.

The wider FOVs afforded the opportunity to extend the pointing duration and extract longer orbital arcs for targets of interest while still covering a large area of sky.
For each SkyMapper pointing, a sequence of 32$\times$10\,s exposures was captured, sandwiched by two 1\,s exposures.
The 1\,s exposures served as a backup source of accurate astrometry in case our automated routines struggled to fit the $\sim$300\,px-long star trails, although the method outlined in Section~\ref{sec:astrometric-calibration} ultimately proved robust.
SkyMapper exposures are separated by 22\,s of dead time (comprising readout and other overheads), so an exposure time of 10\,s was once again chosen to strike a balance between maximising duty cycle and minimising star streak coverage within the survey frames.
A similar duty cycle for the Bisei 1\,m was achieved with a shorter exposure time of 4\,s, in keeping with the instrument's normal survey operations.
The Bisei observations were scheduled to take place contemporaneously with those of SkyMapper, so that both instruments would observe the same field for each pointing.

\begin{table}[tbp]
	\renewcommand{\arraystretch}{1.2}
	\centering
	\caption{Specifications for the SkyMapper, Bisei 1\,m and CLASP\textsuperscript{a} instruments employed for the follow-up observation campaign introduced in Section~\ref{sec:future-work}.}
	\label{tab:instrument-specification-followup}
	\begin{tabular}{@{}llll@{}}
        \toprule
             & SkyMapper & Bisei 1\,m & CLASP \\
        \midrule
            Aperture [m] & 1.35 & 1.00 & 0.36 \\
            Focal ratio & f/4.8 & f/3.0 & f/2.2 \\
            Format [px] & (32$\times$)2k$\times$4k & (4$\times$)2k$\times$4k & 9600~$\times$~6422 \\
            FOV & $2.37^\circ\times2.39^\circ$ & $2.40^\circ\times1.20^\circ$ & $2.63^\circ\times1.76^\circ$ \\
            Pixel scale [$''$px$^{-1}$] & 0.5 & 1.0 & 1.0 \\
            Readout time [s] & 22 & 13 & Negligible\textsuperscript{b} \\
            Exposure time [s] & 10 & 4 & 10 \\
            Filter & Clear\textsuperscript{c} & $Wi$~\cite{okumura2012wide} & Clear\textsuperscript{c} \\
        \botrule
	\end{tabular}
    \footnotetext{\textsuperscript{a}~Note that we provide details for one of the twin instruments comprising CLASP at the time of the campaign.}
    \footnotetext{\textsuperscript{b}~The CLASP instrument's QHY 600M CMOS detector uses a rolling electronic shutter, achieving negligible deadtime due to readout between exposures, but introducing a time offset between each row of the detector.}
    \footnotetext{\textsuperscript{c}~Note that the wavelength sensitivities of the SkyMapper CCD and CLASP CMOS detectors differ, with the latter peaking more strongly in the blue and the former with fairly comprehensive sensitivity across the visible spectrum~\cite{onken2024skymapper}.}
\end{table}

Stars would drift $\sim4.5^\circ$ during the 18-minute pointing, so fields were selected sufficiently far away from the penumbral edge of the Earth's shadow to avoid drifting into eclipse.
At the end of each exposure sequence, the telescope would slew to update its pointing, and a new sequence would begin. 
By the end of the night, this process would cover a strip of sky along, above, or below the geostationary belt, though some gaps in coverage would remain due to the significant drift of the target field between pointings.

\begin{figure}[tbp]
\centering
\includegraphics[width=\textwidth]{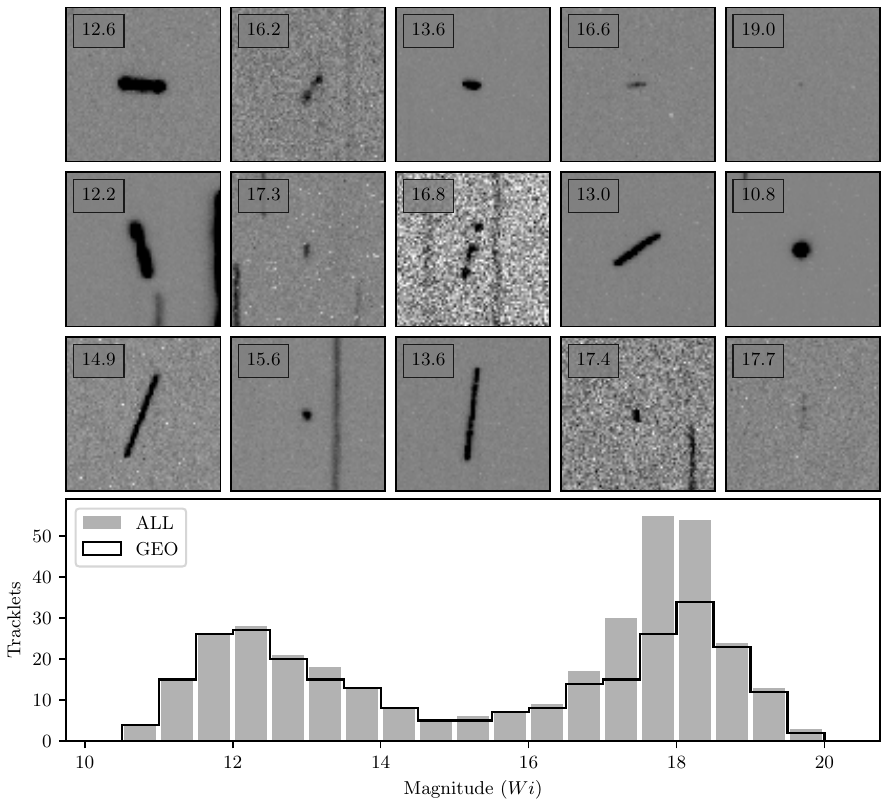}
\caption{The top three rows provide examples of target morphologies extracted from the SkyMapper dataset.
Each 100\,px\,$\times$\,100\,px panel contains a detection associated with a separate tracklet extracted from the SkyMapper survey frames. 
Labels indicate the calibrated magnitude of the detection.
For clarity, we have binned the images by a factor of 2.
Vertical features are stars streaking through the images.
The bottom panel shows a preliminary histogram of median calibrated brightness measurements for the tracklets detected by the Bisei 1\,m telescope, during 59\,hr of observing time between August 2022 and January 2023.
Detections that survive the circular GSO rate cuts defined in Fig.~\ref{fig:int-brightness-histograms} are indicated by the solid black line, labelled `GEO'.
Note that multiple tracklets may be associated with the same RSO; an association between the pointings will be conducted as the survey analysis progresses.}
\label{fig:skymapper-bisei-detections}
\end{figure}

While the collecting areas of SkyMapper and the Bisei 1\,m are much smaller than that of the INT, the instruments were still able to probe the faint end of the bimodal brightness distribution.
In Fig.~\ref{fig:skymapper-bisei-detections}, we present preliminary findings from both the SkyMapper and Bisei datasets.
The left-hand panel shows a histogram of median brightness measurements for tracklets extracted from the Bisei survey frames.
As for Fig.~\ref{fig:int-brightness-histograms}, the calibrated magnitudes here are corrected for the RSO range and Sun distance at the time of the observation. 
However, it is important to note that the two distributions cannot be directly compared, as the Bisei observations were carried out using the wide-band $Wi$ filter~\cite{okumura2012wide}, `redder' than the Harris $V$ filter used for the INT survey, thus introducing a colour offset that will be unique to each detected RSO~\cite[see, e.g.,][]{airey2025comprehensive}.
It is also important to bear in mind that the tracklets have yet to be correlated between pointings, so multiple tracklets may currently be registered for the same RSO if observed across multiple pointings.
In particular, we anticipate a reduction in the bright end of the distribution, as the geostationary belt was observed on multiple nights, and it is likely that station-kept satellites were detected on more than one occasion.
The two distributions are nevertheless very similar, with the bright ends centred on 12$^\text{th}$ magnitude and faint ends beginning to rise at roughly 17$^\text{th}$ magnitude, before dropping off with the sensitivity of the instrument.

Alongside the Bisei histogram, we provide a montage of SkyMapper detections, illustrating the wide variety of target morphologies extracted from the survey frames.
Many of the fainter targets shown were detected towards the end of the night when SkyMapper was pointing in a westerly direction.
Siding Spring and Bisei are more favourably located to probe the population of debris librating about the geopotential well at 75$^\circ$\,E above the Himalayas, which lies below the horizon from the vantage point of La Palma.
While the elevation angle of the well centre was below the instrument limits, we were able to survey the ingress of the well in the latter stages of the night.

Already from the trail morphologies, we see evidence of significant brightness variability. 
The small 0.5$''$px$^{-1}$ pixel scale of the SkyMapper frames, coupled with the 10\,s exposure time, will allow for similarly high-cadence light curves as those presented in Section~\ref{sec:brightness-variability} to be extracted for trailing detections.
The instrument's large readout time, however, will once again result in large gaps in the light curves.

With the CLASP dataset, we plan to explore the benefits of CMOS technology when tasked with monitoring the brightness of RSOs.
At the time of the survey, CLASP featured two 36\,cm RASAs mounted on a Planewave L-600 direct-drive mount, paired with QHY600M CMOS detectors.
Scientific-grade CMOS devices are now available in large formats, with fast, low-noise readout electronics and comparable quantum efficiencies to their CCD counterparts~\cite{alarcon2023scientific}.
When combined with the affordability of mass-produced, small-aperture telescopes, CMOS detectors have the potential to revolutionise wide-field surveillance for both astronomy~\cite[see, e.g.,][]{law2022low} and space domain awareness~\cite[see, e.g.,][]{blake2023exploring}, offering high-cadence imaging of large areas of sky with impressive depth, especially when combined with the stacking technique discussed in Section~\ref{sec:blind-stacking}.
The negligible readout time for CLASP will enable the extraction of high-cadence light curves with no gaps between frames, providing a near-continuous view of targets during each pointing of the telescope.

For the CLASP observations, we opted to carry forward the observational strategy used for SkyMapper, including the exposure time of 10\,s, as opposed to imaging at the high frame rates achievable with the CMOS detector.
For sufficiently bright RSOs, this approach offers a way to avoid the high data rates and volumes associated with high-cadence imaging, while still benefitting from the fast readout of the CMOS detectors.
We are also exploring neuromorphic camera technology~\cite{cohen2019event} as an alternative solution.
Inspired by the biological retina, event-based sensors output asynchronous streams of events instead of frames of pixel intensities. 
An `event' is generated only when a pixel detects a change in log light intensity, thus enabling low data rates relative to frame-based counterparts.

%We highlight the following areas of future work associated with the presented study \comment{TODO}:
%\begin{itemize}
    %\item Apply more sophisticated short arc determination techniques to the orbital arcs.
    %\item Work is underway to analyse the full SkyMapper and Bisei datasets from the follow-up observation campaign, totalling 239\,hr and 59\,hr of observation time, respectively.
    %With the contemporaneous pointings, we aim to exploit the parallax effect to obtain more accurate orbits for the extracted arcs.
    %\item Analysis of the CLASP observations is underway.
    %We aim to extract high-cadence, gapless light curves for graveyard RSOs captured by the instrument, exploiting the beneficial characteristics of scientific-grade CMOS technology.
%\end{itemize}

%\begin{table}[h]
%\caption{Caption text}\label{tab1}%
%\begin{tabular}{@{}llll@{}}
%\toprule
%Column 1 & Column 2  & Column 3 & Column 4\\
%\midrule
%row 1    & data 1   & data 2  & data 3  \\
%row 2    & data 4   & data 5\footnotemark[1]  & data 6  \\
%row 3    & data 7   & data 8  & data 9\footnotemark[2]  \\
%\botrule
%\end{tabular}
%\footnotetext{Source: This is an example of table footnote. This is an example of table footnote.}
%\footnotetext[1]{Example for a first table footnote. This is an example of table footnote.}
%\footnotetext[2]{Example for a second table footnote. This is an example of table footnote.}
%\end{table}

\section{Conclusions}
\label{sec:conclusions}

In this work, we have presented a reanalysis of images from a survey of faint geosynchronous debris conducted with the 2.54\,m Isaac Newton Telescope (INT) at the Roque de los Muchachos Observatory, La Palma.
A combination of improved star trail centroiding and iterative distortion fitting has significantly enhanced the astrometric calibration stage of the processing pipeline, achieving sub-arcsecond accuracies.
We have boosted target recovery by applying a multi-frame blind stacking technique, unearthing 25 tracklets previously missed by single-frame extraction methods and pushing the survey's sensitivity limit fainter by 1 magnitude.
The stacking pipeline has also been applied to a contemporaneous dataset acquired with a commercial-off-the-shelf (COTS) 36\,cm robotic astrograph (RASA), offering a unique opportunity to benchmark the performance of the algorithm through the attempted recovery of INT detections from the COTS survey frames.
After accounting for rate limits constraining the blind stacking search, we found that the technique was able to recover all feasible INT targets brighter than $V\sim17.5$ from the RASA images.
Furthermore, the RASA instrument's superior duty cycle made the COTS dataset far better suited for blind stacking; the pipeline was able to recover 62 tracklets missed by the original single-frame approach, pushing the sensitivity limit fainter by roughly 2 magnitudes.

With the improved astrometry, we were able to perform an initial orbit determination for the very short INT arcs.
The yielded solutions show clear signs of the 53-year evolutionary cycle in inclination and right ascension of the ascending node, undertaken by faint, uncontrolled fragments of debris.
Future efforts will seek to apply more sophisticated and robust approaches to improve convergence for shorter arcs.
We also carried out a quantitative analysis of the brightness variability exhibited by the trailing detections extracted from the survey frames.
The faint end of the sampled population was found to be proportionally more variable, with over 10\% of light curves showing strong signatures of periodicity.

Lastly, we provided an overview of the preliminary findings from a follow-up multinational observation campaign that took place between March 2022 and January 2023, using the 1.35\,m SkyMapper Telescope at Siding Spring Observatory, Australia, the 1\,m telescope at Bisei Space Guard Center, Japan, and the twin 36\,cm Warwick CLASP telescope in La Palma.
Work is underway to analyse the full SkyMapper and Bisei datasets, totalling 259\,hr and 59\,hr of survey time, respectively.
For contemporaneous pointings, we aim to exploit the parallax effect to obtain higher quality orbit solutions for shared debris detections.
In addition, the rapid readout of the CLASP instrument will enable the extraction of high-cadence, gapless light curves for the observed targets.

With space traffic management concerns beginning to extend beyond GSO altitudes, the need to search deeper and uncover fainter RSOs with high-sensitivity measurements has never been greater. 
Scientifically driven surveys of high-altitude orbits have an important role to play in understanding the nature and evolution of the optically faint debris environment.
More extensive and persistent observation campaigns supported by large-aperture telescopes are needed to maintain custody of very faint GSO debris, build durable catalogues, and further characterise their behaviour to more comprehensively assess the risk they pose to operational satellites in the region.

\backmatter

%\bmhead{Supplementary information}

%If your article has accompanying supplementary file/s please state so here. 

%Authors reporting data from electrophoretic gels and blots should supply the full unprocessed scans for key as part of their Supplementary information. This may be requested by the editorial team/s if it is missing.

%Please refer to Journal-level guidance for any specific requirements.

%\bmhead{Acknowledgements}

%Acknowledgements are not compulsory. Where included they should be brief. Grant or contribution numbers may be acknowledged.

%Please refer to Journal-level guidance for any specific requirements.

\section*{Declarations}

%Some journals require declarations to be submitted in a standardised format. Please check the Instructions for Authors of the journal to which you are submitting to see if you need to complete this section. If yes, your manuscript must contain the following sections under the heading `Declarations':

%\begin{itemize}
%\item Conflict of interest/Competing interests (check journal-specific guidelines for which heading to use)
%\item Ethics approval and consent to participate
%\item Consent for publication
%\item Data availability 
%\end{itemize}

%\noindent
%If any of the sections are not relevant to your manuscript, please include the heading and write `Not applicable' for that section. 

\subsection*{Funding}

JAB acknowledges support from the Science and Technology Facilities Council (grant ST/Y50998X/1).
BC acknowledges support from the Defence Science and Technology Laboratory (UK).
This paper uses data from the following instruments: the Isaac Newton Telescope, operated by the Isaac Newton Group of Telescopes, and the CLASP telescope, operated by the University of Warwick, on the island of La Palma, at the Spanish Observatorio del Roque de los Muchachos of the Instituto de Astrofisica de Canarias; the SkyMapper Telescope, operated at Siding Spring Observatory, Coonabarabran, by the Australian National University; and the 1\,m telescope at Bisei Space Guard Center, operated by the Japan Aerospace Exploration Agency.

\subsection*{Competing interests}

The authors have no competing interests to declare that are relevant to the content of this article.

\subsection*{Author contribution}

JAB wrote the manuscript, developed the upgraded astrometric calibration procedure, and carried out the light curve and brightness variability analyses.
BC and CP contributed to the writing of the manuscript. 
BC developed the adapted blind stacking pipeline, and CP performed the initial orbit analyses.
CP, RA, DV, SE, and TS provided support in interpreting the initial orbit results.
RA, IA, LB, AMC, ISL, AM, JM, MAM, BS, and PW provided insights that strengthened discussion throughout the manuscript.
DP, JAB, PC, DV, and SE compiled the observing proposal that was awarded 8 nights of dark-grey time on the INT.
JAB, PC, GP, and WF carried out the INT observations. 
PC obtained the contemporaneous RASA dataset.
For the follow-up observation campaign, CO and CW carried out the SkyMapper observations, while TF, DK, TN, KN, SiO, and SU acquired the contemporaneous Bisei dataset.
BC obtained the CLASP observations.
GP developed the scheduler for the SkyMapper-Bisei observations.
JAB developed the processing pipeline for the SkyMapper data, while TY processed the Bisei frames.
JAB, BC and CP prepared the figures.
All authors reviewed the manuscript.

%\bigskip

%\bigskip\noindent
%Springer journals and proceedings: \url{https://www.springer.com/gp/editorial-policies}

%\bigskip\noindent
%Nature Portfolio journals: \url{https://www.nature.com/nature-research/editorial-policies}

%\bigskip\noindent
%\textit{Scientific Reports}: \url{https://www.nature.com/srep/journal-policies/editorial-policies}

%\bigskip\noindent
%BMC journals: \url{https://www.biomedcentral.com/getpublished/editorial-policies}
%\end{flushleft}

\begin{appendices}

\section{RASA astrometry}
\label{app:rasa-astrometry}

%An appendix contains supplementary information that is not an essential part of the text itself but which may be helpful in providing a more comprehensive understanding of the research problem or it is information that is too cumbersome to be included in the body of the paper.

For ease of reference, we provide plots of astrometric accuracy for the RASA survey frames in Fig.~\ref{fig:astrometry-rasa}, analogous to those shown in Fig.~\ref{fig:astrometry-int-chips} for the INT dataset, before and after applying the iterative fitting approach outlined in Section~\ref{sec:astrometric-calibration}. 
As for the INT frames, we see a higher accuracy in the cross-trail (declination) direction compared to the along-trail (right ascension) direction, owing to the greater extent of the trail along the longitudinal axis.
The slightly lower quality for the bottom-left corners of the frames is due to PSF degradation caused by miscollimated optics, an issue that was corrected following the survey.

\setcounter{figure}{18}
\renewcommand{\thefigure}{\arabic{figure}}

\begin{figure}[tbp]
\centering
\includegraphics[width=\textwidth]{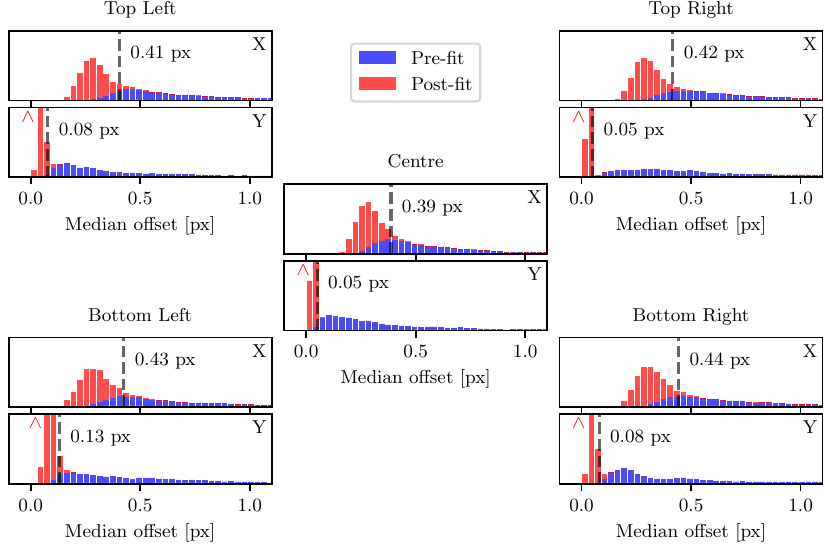}
\caption{Pre- (blue) and post-fit (red) measures of the astrometric accuracy for the RASA survey frames.
We plot the median offsets in right ascension-declination space between the extracted star trail positions and their matched comparison stars.
The top panels of each region show median offsets in the X (right ascension) direction, while those in the bottom panels correspond to the Y (declination) direction.
We consider matched stars within a 2044\,px$\times$2044\,px tile situated in the top left, top right, centre, bottom left and bottom right regions of a given frame, as labelled.
In each panel, we label the 95$^\text{th}$ percentile achieved by the iterative fitting algorithm.
Adapted from \cite{blake2020supplementing}.}
\label{fig:astrometry-rasa}
\end{figure}

\section{Updated INT photometry}
\label{app:updated-int-photometry}

As part of the iterative fitting stage of the upgraded astrometric calibration procedure outlined in Section~\ref{sec:astrometric-calibration}, we generate new estimates of the photometric zero points for the INT survey frames.
In the middle panels of Fig.~\ref{fig:int-detector-stats}, we show the distribution of detections in XY space across the three CCD chips comprising the WFC mosaic.
We observe a relatively even spread spatially, though with fewer detections stemming from CCD 3, likely due to strong vignetting, which resulted in a slightly inflated background fluctuation, even after flat-field correction.
The mean zero point remains consistent at $V\sim25.2$ for all three chips.
The brightness measurements for INT detections presented in Sections \ref{sec:blind-stacking} and \ref{sec:brightness-variability} account for the updated photometry.

\setcounter{figure}{19}

\begin{figure}[tbp]
\centering
\includegraphics[width=\textwidth]{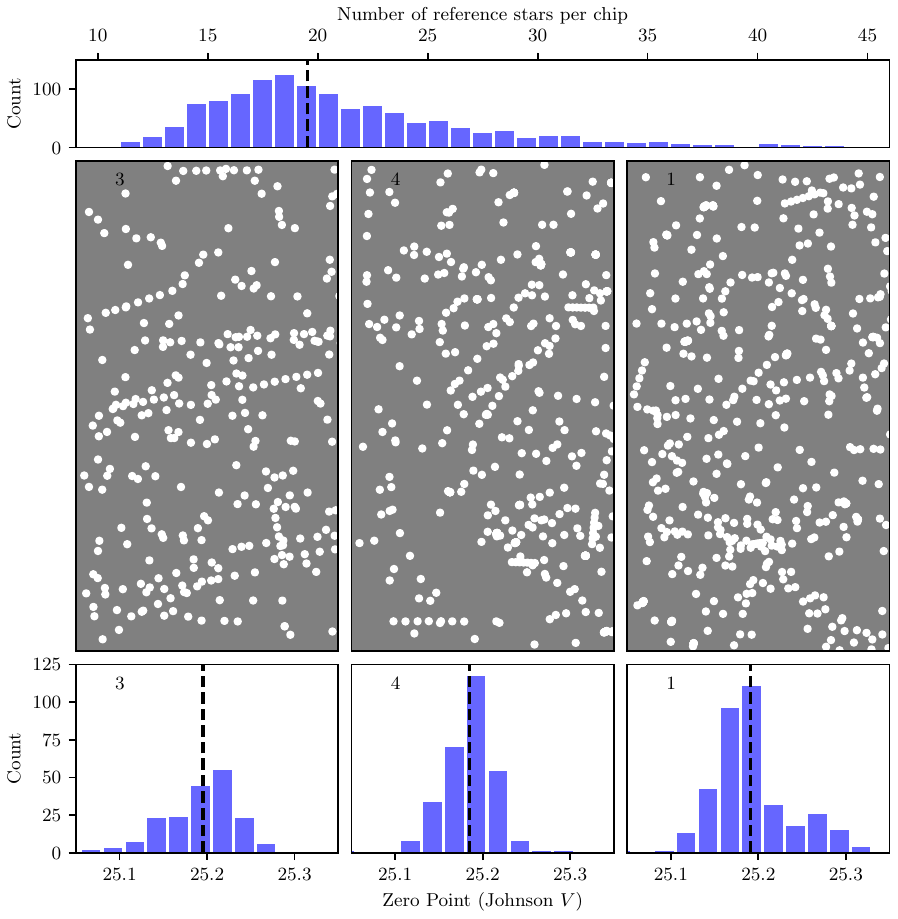}
\caption{Photometric results for the INT survey frames associated with detections.
The top panel shows the number of comparison stars extracted per CCD chip for the frames.
The middle panels, labelled by chip ID, show the central positions of detections (white) across the detector (grey). 
Below the detector panels, we provide updated photometric zero point distributions for the three chips. 
Black dashed lines indicate the mean value in each histogram.}
\label{fig:int-detector-stats}
\end{figure}

\section{Blind stacking examples}
\label{app:blind-stacking-examples}

In Fig.~\ref{fig:example-stacking-1408411} and Fig.~\ref{fig:example-stacking-1407379}, we provide additional examples of new INT detections unearthed by the adapted blind stacking pipeline introduced in Section~\ref{sec:blind-stacking-technique}.
As before, the targets are barely visible, if at all, in the individual frames, but their SNR is vastly improved by the stacking and integration steps of the pipeline.
The flux ranges extracted for the new INT detections are given as a function of angular rate in Fig.~\ref{fig:int-stacking-flux}.
For faster moving targets, we see that the minimum flux is typically higher, owing to the sensitivity of the technique; the spreading of the flux over more pixels will result in a lower SNR, heightening the impact of any uncertainty associated with pixel alignment and blending.

\setcounter{figure}{20}

\begin{figure}[tbp]
\centering
\includegraphics[width=\textwidth]{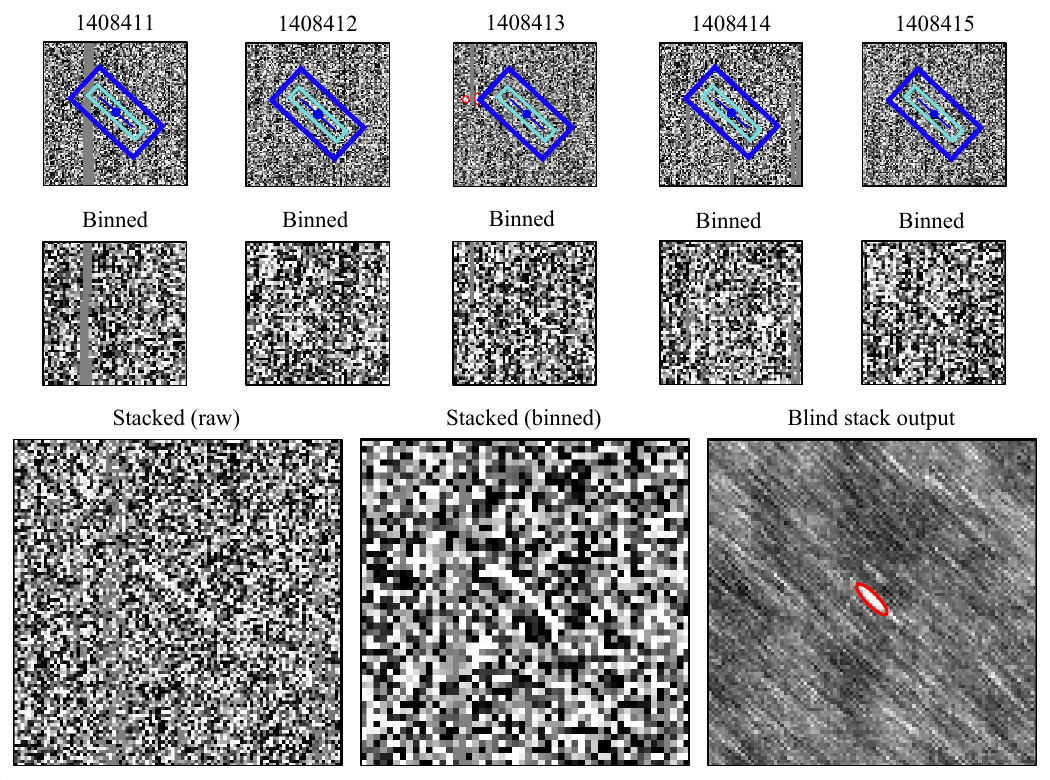}
\caption{An example recovery of a new INT target, with range-corrected brightness $V\sim22.5$, found using the adapted blind stacking pipeline.
The top row shows five survey frames, while the middle row gives their ($2\times2$) binned counterparts.
From left to right on the bottom row, we show the median stacks of the raw and binned survey frames, alongside the output of the stacking pipeline.
Blue boxes highlight the target location in the individual frames with the blue point and line showing the predicted extent of the detection and the cyan box showing the centroid corrected rectangular aperture. The red ellipse shows the \texttt{SEP} detection in the stacking output frame.}
\label{fig:example-stacking-1408411}
\end{figure}

\setcounter{figure}{21}

\begin{figure}[tbp]
\centering
\includegraphics[width=\textwidth]{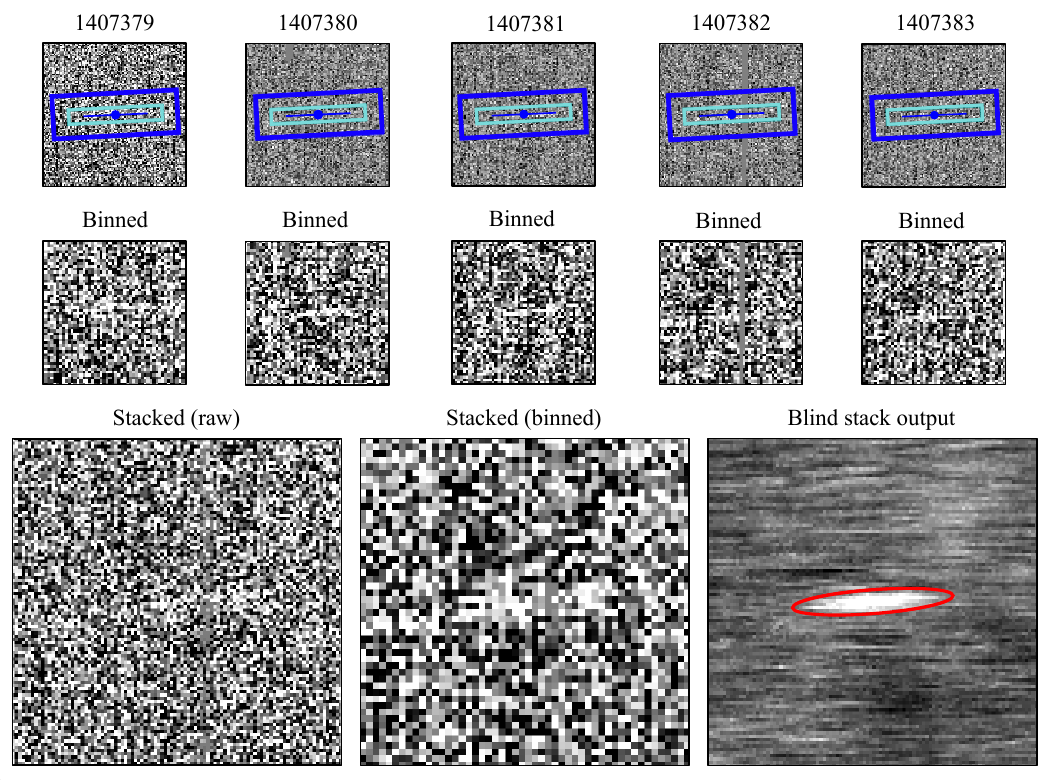}
\caption{An example recovery of a new INT target, with range-corrected brightness $V\sim20.5$, found using the adapted blind stacking pipeline.
The top row shows five survey frames, while the middle row gives their ($2\times2$) binned counterparts.
From left to right on the bottom row, we show the median stacks of the raw and binned survey frames, alongside the output of the stacking pipeline.
Blue boxes highlight the target location in the individual frames with the blue point and line showing the predicted extent of the detection and the cyan box showing the centroid corrected rectangular aperture. The red ellipse shows the \texttt{SEP} detection in the stacking output frame.}
\label{fig:example-stacking-1407379}
\end{figure}

\setcounter{figure}{22}

\begin{figure}[tbp]
\centering
\includegraphics[width=\textwidth]{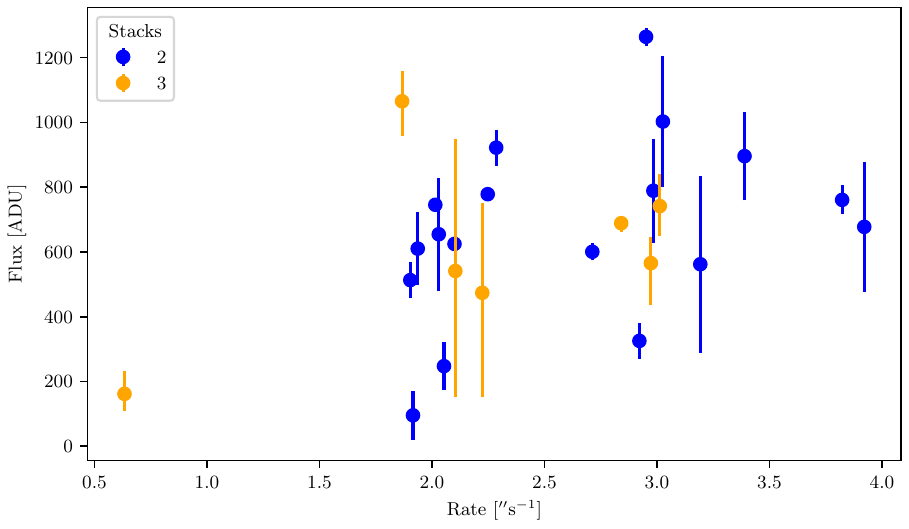}
\caption{Extracted flux measurements for the 25 targets found by applying the adapted blind stacking algorithm (see Section~\ref{sec:blind-stacking-technique}) to the INT survey frames, plotted as a function of angular rate.
The central value gives the median flux, while the error bars span the range of fluxes across the observing window.
For some targets, the error bars are too small to be seen.}
\label{fig:int-stacking-flux}
\end{figure}




\end{appendices}


\bibliography{sn-bibliography}% common bib file

\end{document}